\documentclass[10pt]{article}

\usepackage{graphicx}
\graphicspath{{converted_graphics/}}
\begin{document}

\title{Cosmological model with variable equations of state for matter and dark energy}
\author{J. Ponce de Leon\thanks{E-Mail:
jpdel1@hotmail.com}  \\Laboratory of Theoretical Physics, 
Department of Physics\\ 
University of Puerto Rico, P.O. Box 23343,  
San Juan,\\ PR 00931, USA}
\date{March, 2012}

\maketitle
\begin{abstract} 
We construct  a cosmological model which is physically reasonable, mathematically tractable, and extends the study of CDM models to the case where the equations of state (EoS) for matter and dark energy (DE) vary with time. It is based on the assumptions of (i) flatness, (ii) validity of general relativity, (iii) the presence of a DE  component that varies between two asymptotic values,  
(iv) the  matter of the universe smoothly evolves from an initial radiation stage - or a barotropic perfect fluid - to a phase where it behaves as cosmological dust at late times. The model approximates the CDM ones for small $z$ but significantly differ from them for large $z$. We focus  our attention on how the evolving EoS for matter and DE can modify the CDM paradigm. 
We discuss  a number of physical scenarios. 
One of them includes, as a particular case, the so-called generalized Chaplygin gas models where DE evolves from non-relativistic dust. Another kind of models shows that the current accelerated expansion is compatible with a DE  that behaves like pressureless dust at late times.   We also find that a universe with variable DE can go from decelerated to accelerated expansion, and vice versa, several times.

\end{abstract}

\bigskip

PACS numbers: 98.80.Es, 98.80.-k, 95.36.+x, 98.80.Cq, 04.20.-q

Keywords: Cosmological constant, Cosmology,   Einstein equation, general relativity, Dark energy,  Cosmic Late-time Acceleration

\newpage
\section{Introduction}

One of the most challenging problems in cosmology today is to explain the observed late-time accelerated expansion of the universe \cite{Riess}-\cite{Sievers}. 
Since the gravity of both baryonic (ordinary) matter and radiation is attractive, the fact that the universe is presently
accelerating, and may continue to do so, forces us to rethink and question some fundamental concepts about the universe.\footnote{It should be mentioned that some authors keep a more skeptical  point of view. They argue that the observational data,  as it presently stands,  can be explained  without resorting to the existence of a negative-pressure fluid or a cosmological constant. The concept is that the departure of the observed universe from an Einstein-de Sitter model can be ascribed to other physical processes  and/or to the influence of inhomogeneities. See e.g., \cite{Blanchard},  \cite{Alnes}, \cite{Marie}  and references therein.}

The first question highlights our limited knowledge of the real nature of the content of the universe. In fact, an accelerated expansion requires the presence  of a new form of matter (called {\it dark energy}), which could (i) produce gravitational repulsion, i.e., violate the strong energy condition; (ii) account for  $70\%$ of the total content of the universe; (iii) remain unclustered on all scales where gravitational clustering of ordinary matter  is seen. (For a recent review see Ref. \cite{Review}).
While we do not yet know exactly the physical mechanism responsible for
this exotic behavior, we do have  
possible candidates for dark energy (DE). They include: a cosmological constant or a time dependent cosmological term \cite{Peebles}-\cite{Padmanabhan0};  an evolving scalar field known as {\it quintessence} ($Q$-matter) with a potential  giving rise to negative pressure at the present epoch \cite{Zlatev}-\cite{Deustua}; dissipative fluids \cite{DissipativeFluids}; Chaplygin gas \cite{Chaplygin gas 1}-\cite{Chaplygin gas 2}; K-essence \cite{K-essence 1}-\cite{K-essence 4}, and other more exotic models \cite{domain walls}.

The second question is whether  general relativity is applicable to describe the universe as a whole. Indeed, this and other puzzles of theoretical and experimental gravity have triggered a huge  interest in alternative theories of gravity (See, e.g. \cite{Gilles} and references therein) where the cosmological acceleration is
not provided by dark energy, but rather by a modification of the Friedman equation on large scales \cite{Defayet}, \cite{Dvali}. These include scalar-tensor theories of gravity \cite{Bertolami}-\cite{Chakraborty}, various versions of Kaluza-Klein theories, braneworld, STM and Brans-Dicke theory  in $5D$ \cite{Das2}-\cite{JPdeL}.

In view of these uncertainties, much of the research work being done in cosmology is based on the construction and study of specific cosmological models. The simplest one that predicts accelerated cosmic expansion and fits observational data reasonably well  is the $\Lambda$CDM  model 
\cite{Riess}-\cite{Riess2}  which is
based on the assumptions of (i) flatness, (ii) validity of general relativity, (iii) the presence of a cosmological constant $\Lambda$ and
(iv) Cold Dark Matter (CDM). The main problem of this model is the huge difference between the observed value of the cosmological constant and the one predicted in quantum field theory. 
The other one, although not vital for the model, is that the assumption of CDM can not be applied to the entire evolution of the universe.

In this work, we construct  an alternative  cosmological model where we keep the first two assumptions but relax the other two. Rather than
choosing to investigate constraints on specific DE models, here we  use a  parameterization originally proposed by Hannestad and  M\"{o}rtsell \cite{EoS} which can accommodate a number of  DE models, including a cosmological constant. On the other hand, we employ a phenomenological approach to describe the  matter content of the universe as 
a mixture of different components,  which are not required to expand adiabatically. The mixture smoothly evolves from an initial dense radiation stage - or a barotropic perfect fluid - to a phase where it behaves as cosmological dust at late times.  

This framework allows us to compare and contrast different physical settings and tackle a number of questions.  Here we  focus our attention on the evolution of the universe and its cosmic acceleration  and on how the evolving EoS for matter and DE might modify the $\Lambda$CDM paradigm.
We explore whether DE could play a significant role in the past
evolution of our universe, and - conversely - whether the primordial EoS of matter can affect the details of accelerated expansion at late times. 
Also,  whether   dark energy can cross the line between quintessence and phantom regimes (the crossing of the cosmological constant boundary) \cite{stefancic}.  
Another important question is whether the accelerated expansion, once begun, continues forever: Is it possible for an ever-expanding DE-dominated universe to go through different cycles in which it changes from decelerated to accelerated expansion, and vice versa?

This paper is organized as follows. In section $2$ we give a brief introduction to Einstein's equations in a homogeneous and isotropic background. In section $3$ for the sake of generality we integrate the field equations without assuming any specific EoS.  In this way we obtain general expressions for the Hubble, density and deceleration parameters which in practice provide a simple recipe for the construction of cosmological models.
In section $4$ we introduce the EoS that generate our cosmological model and obtain explicit expressions for the relevant cosmological quantities.  In section $5$ we study the properties of a universe whose matter content is described by the EoS proposed in section $3$, and the DE has a constant EoS.    
In section $6$, in the context of CDM, we concentrate our study to the effects of a DE with variable EoS. In section $7$ we present a summary of our work.

\section{Field equations}

 A spatially homogeneous and isotropic universe is described by the FLRW line element, which in polar coordinates $(r, \theta, \varphi)$ is written as\footnote{In this work we use relativistic units: $8\pi G = c = 1$. Also a subscript zero  denotes the quantities given  at the current epoch.}
\begin{equation}\label{2.1}
ds^{2}= dt^{2} - a^{2}(t)\left[\frac{dr^{2}}{1 - k r^2}+r^{2}(d\theta^{2}+{\sin}^{2}\theta
d\varphi^{2})\right],
\end{equation}
where $a(t)$ is the scale factor with cosmic time $t$; $k$ is the curvature signature which can, by a suitable scaling of $r$, be set equal to $- 1$, $0$ 
or $ + 1$.

The evolution of the scale factor and the stress energy in the universe are  governed by the Einstein field equations:
\begin{equation}\label{2.2}
R_{\mu\nu}-\frac{1}{2} g_{\mu\nu}R = T_{\mu\nu}.
\end{equation}
The assumption of isotropy and homogeneity (\ref{2.1}) requires that the stress-energy tensor take on the perfect fluid form\footnote{To be rigorous, one should say that the matter has to be a fluid with bulk viscosity at most.}: $T_{\nu}^{\mu} =  \mbox{diagonal}\left(\rho, - p, - p, - p\right)$, where $\rho$ and $p$ stand for the {\it{total}} energy density and  {\it{total}} isotropic pressure of the cosmological ``fluid". This  framework  leads to two independent equations 
\begin{equation}
\label{field equations 1}
\frac{3{\dot{a}}^2}{a^2} + \frac{3 k}{a^2} = \rho =  \sum_{i = 1}^{N}\rho^{(i)},
\end{equation}
and
\begin{equation}
\label{field equations 2}
\frac{2\ddot{a}}{a} + \frac{{\dot{a}}^2}{a^2} + \frac{k}{a^2} = - p = - \sum_{i = 1}^{N} p^{(i)}, 
\end{equation}
where an over-dot indicates ordinary derivative with respect to $t$.  The total energy density and pressure have been split up into constituents: $\rho^{(i)}$ and $p^{(i)}$ represent the energy density and pressure of the $i$-th component that fills the universe; the sums are over all, say $N$, different  species of matter  present in the universe (baryonic, non-baryonic, 
 radiation, cosmic neutrinos, dark energy, etc.) at a given epoch. 

These equations can be combined to obtain the continuity equation
\begin{equation}
\label{continuity equation}
\dot{\rho} + 3\left(\rho + p\right)\frac{\dot{a}}{a} = 0,
\end{equation}
which is equivalent to the covariant conservation equation $T^{\mu\nu}_{\;\;\;;\nu} = 0$.

Thus, there are two independent equations and  $(2N + 1)$ unknown quantities, namely, $a(t)$,  $\rho^{(i)}$, $p^{(i)}$, $i = 1..N$. To close the system we need to provide $(2N - 1)$ additional equations. 
They could be $N$ equations of state (EoS) relating the pressures and densities. The remaining $\left(N - 1\right)$ equations are usually  generated  by the assumption that each component is expanding adiabatically, i.e., that there is no interaction between the cosmological constituents.  What this means is that one imposes the energy conservation equation (\ref{continuity equation}) on each constituent.

\medskip

The scale factor $a(t)$ can be expressed as a Taylor series around the present time $t_{0}$:  

\begin{equation}
\label{expansion of a}
\frac{a(t)}{a_{0}} = 1 + H_{0}\left(t - t_{0}\right) - \frac{1}{2}q_{0}H_{0}^2 \left(t - t_{0}\right)^2 + \frac{1}{6}r_{0}H_{0}^3 \left(t - t_{0}\right)^3 + ...,
\end{equation}
where $H = \dot{a}/a$ is the Hubble parameter, which is given directly by  (\ref{field equations 1}); $q = - \ddot{a} a/\dot{a}^2$ is the deceleration parameter that can be evaluated from (\ref{field equations 2}). If the $i$-th component has an EoS $p^{(i)} = w^{(i)}\rho^{(i)}$ then

\begin{equation}\label{most general expression for q}
q = \frac{1}{2}\left[1 + 3 \sum_{i = 1}^{N}w^{(i)} \Omega^{(i)}\right] + \frac{k}{2 a^2 H^2},
\end{equation}
where $\Omega^{(i)} = \rho^{(i)}/3 H^2$ is the corresponding density parameter. Next, $r = \stackrel{\cdots}{a}/a H$ is the statefinder parameter introduced by Sahni {\it{et al.}} \cite{Sahni} and Ulam {\it{et al.}} \cite{Ulam}. Taking time derivative in (\ref{field equations 2}) we obtain
\[
r = 1 - \frac{1}{2 H^3}\sum_{i = 1}^{N}\left[{\dot{w}}^{(i)}\rho^{(i)} + w^{(i)}{\dot{\rho}}^{(i)}\right] + \frac{k}{a^2 H^2}.
\]
If, for the sake of generality,  the interplay between the constituents is not neglected then each one evolves as 
\[
{\dot{\rho}}^{(i)} + 3 H \left[1 + w^{(i)}\right]\rho^{(i)} = Q^{(i)}, \;\;\;\mbox{with}\;\;\;\sum_{i = 1}^{N}Q^{(i)} = 0,
\]
where 
$Q^{(i)}$ measures the strength of the interaction of the $i$-th constituent with the rest of the components. Using this equation, the parameter $r$ becomes

\begin{equation}
\label{general expression for r}
r = 1 + \frac{3}{2}\sum_{i = 1}^{N}\left\{3 w^{(i)}\left[1 + w^{(i)}\right] - a\, \frac{d w^{(i)}}{d a}\right\} \Omega^{(i)} - \frac{1}{2 H^3}\sum_{i = 1}^{N}w^{(i)}Q^{(i)}+ \frac{k}{a ^2 H^2}.
\end{equation}
In a similar way, one can express higher order terms in  (\ref{expansion of a}) as functions of $w^{(i)}$, $Q^{(i)}$ and their derivatives.

\section{General integration of the field equations}

The purpose of this section is to integrate the field equations with the least possible number of assumptions.
With this in mind,    it is convenient  to split up  the total energy density and pressure as 
\begin{eqnarray}
\label{separation of density and pressure}
\rho &=& \rho^{(DE)} + \rho^{(M)},\nonumber \\
p &=& p^{(DE)} + p^{(M)},
\end{eqnarray}
where $\rho^{(DE)}$ and $p^{(DE)}$ represent the DE contribution and  
\begin{equation}
\label{definition of rhom and pm}
\rho^{(M)} = \sum_{i}\rho^{(i)}, \;\;\;p^{(M)} = \sum_{i} p^{(i)},
\end{equation}
are hybrids containing the contribution of relativistic particles, photons, the three neutrino species, as well as  the contribution of non-relativistic particles (baryons, WIMPS, etc.). 

In term of these quantities the field equations (\ref{field equations 1})-(\ref{field equations 2}) now 
contain five unknowns. Here, we formulate no assumptions regarding the nature of the expansion of the constituents of the matter mixture $\rho^{(M)}$; they may evolve non-adiabatically. However, to make the problem solvable, we neglect any matter-DE interaction and assume  that the DE component is expanding adiabatically. As a result,  $\rho^{(DE)}$ and $p^{DE}$,  as well as the effective quantities $\rho^{(M)}$ and $p^{M}$,  satisfy the continuity equation (\ref{continuity equation}) separately.

To close  the system of equations we should provide two EoS.
The simplest equation of state between density and pressure is the so called barotropic equation $p/\rho = w$, where $w$ - in relativistic units - is a dimensionless constant. A direct generalization to this equation is to assume that $w$ is {\it not} a constant but a function of the epoch. In this work we assume that both, matter and DE satisfy such type of EoS, viz.,
\begin{eqnarray}
\label{definition of omegas}
p^{(M)} &=& w(a)\, \rho^{(M)},\nonumber \\
p^{(DE)} &=& W(a)\, \rho^{(DE)}. 
\end{eqnarray}
To explain the accelerated expansion one has to accept that the DE component violates the strong energy condition\footnote{The strong energy condition for perfect fluids, in the comoving frame, requires $\rho + p \geq 0$, $\rho + 3 p \geq 0.$}. Thus, in what follows we assume $W(a) < - 1/3$. 

\medskip

$\bullet$ With these assumptions the field equations (\ref{field equations 1})-(\ref{field equations 2}) become 
\begin{equation}
\label{relation between the Omegas}
\Omega^{(M)} + \Omega^{(DE)} = 1 + \frac{k}{a^2 H^2},
\end{equation}
and
\begin{equation}
\label{general expression for q}
q = \frac{1}{2} + \frac{3}{2}\left[W \Omega^{(DE)} + w \Omega^{(M)}\right] + \frac{k}{2 a^2 H^2}, 
\end{equation}
where  $\Omega^{(M)} = \rho^{(M)}/3 H^2$ and $\Omega^{(DE)} = \rho^{(DE)}/3 H^2$.
These can, formally,  be regarded as two equations for $\Omega^{(M)}$ and $\Omega^{(DE)}$. Solving them we get 
\begin{eqnarray}
\label{general expression for Omega}
\Omega^{(M)}  &=& \frac{2 q - 1 - 3 W}{3 \left(w - W\right)}- \frac{k\left(1 
+ 3 W\right)}{3 a^2 H^2 \left(w - W\right)}, \nonumber \\
\Omega^{(DE)} &=& - \frac{2 q - 1 - 3 w}{3\left(w - W\right)} + \frac{k\left(1 + 3 w\right)}{3 a^2 H^2\left(w - W\right)}. 
\end{eqnarray}
We note that the denominator in these expressions is always positive because $W \leq 0$ for DE, as well as for Chaplygin gas models. Thus, the fact that $\Omega^{(M)} \geq 0$ and $\Omega^{(DE)}\geq 0$ imposes an upper and lower limit on $q$, viz.,
\begin{equation}
\label{limits on q, 1}
\frac{1 + 3 W}{2}\left[1 + \frac{k}{a^2 H^2}\right] \leq q \leq \frac{1 + 3 w}{2}\left[1 + \frac{k}{ a^2 H^2}\right].
\end{equation}
In the epoch where $0 \leq\ \Omega^{(M)} \leq 1/2$ the upper limit reduces to
\begin{equation}
\label{limits on q, 2}
q \leq \frac{1}{2} + \frac{3}{4}\left[w + W\right] + \frac{k}{2 a^2 H^2}\left[1 + 3 w^{(\pm)}\right],
\end{equation}
where $w^{(+)} = W$ if $k = 1$,  and $w^{(-)} = w$ if $k = - 1$.
For the $\Lambda$CDM model\footnote{In the $\Lambda$CDM model the universe is flat, filled with dust $w = 0$ and the DE is attributed to the presence of a cosmological constant $W = - 1$.} the first inequality gives $- 1 \leq  q \leq 1/2$ during the whole evolution and $-1 \leq q \leq - 1/4$ in the DE dominated era.

\medskip

$\bullet$ Given the EoS (\ref{definition of omegas}), the continuity equations for $\rho^{(M)}$ and $\rho^{(DE)}$ can be formally integrated to obtain the evolution of the energy densities as
\begin{eqnarray}
\label{energy densities, in terms of a}
\rho^{(M)} &=& \frac{C_{1}}{a^3}e^{- 3 \int{\frac{w(a)}{a}d a}}, \nonumber \\
\rho^{(DE)} &=& \frac{C_{2}}{a^3}e^{- 3  \int{\frac{W(a)}{a} d a}},
\end{eqnarray}
where $C_{1}$ and $C_{2}$ are constants of integration. Let us use $a_{*}$ to denote the epoch at which $\rho^{(M)}(a_{*}) = \rho^{(DE)}(a_{*})$, and - for algebraic simplicity -  introduce the dimensionless quantity 
\begin{equation}
\label{definition of x}
x \equiv \frac{a}{a_{*}}.
\end{equation}
Using this notation, without loss of generality, we can write
\begin{eqnarray}
\label{energy densities, general}
\rho^{(M)} &=& \frac{C}{x^3}e^{- 3 \int_{1}^{x}{\frac{w(u)}{u}du}}, \nonumber \\
\rho^{(DE)} &=& \frac{C}{x^3}e^{- 3  \int_{1}^{x}{\frac{W(u)}{u} du}},
\end{eqnarray}
where $C$ represents the common value shared by the densities at the transition point\footnote{For non-phantom ($W \geq -1$) and phantom $(W < -1)$ models the equation $\rho^{(M)}(a_{*}) = \rho^{(DE)}(a_{*})$ has no more than one solution, because the DE energy density is strictly decreasing or increasing function of $a$, respectively [$x \frac{d \rho^{(DE)}}{d x} = - 3\left(1 + W\right)\rho^{(DE)}$]. However, this is not necessarily so if the   DE component has different regimes where it crosses the cosmological constant boundary $(W = - 1)$.} $x = 1$ ($a = a_{*}$). Since $W < 0$ and $w \geq 0$, it follows that $\rho^{(M)} > \rho^{(DE)}$ for $x < 1$, and vice versa.

In what follows we denote
\begin{eqnarray}
\label{definition of F and G}
F(x) &=& e^{- 3 \int_{1}^{x}{\frac{w(u)}{u}du}}, \nonumber \\
G(x) &=& e^{- 3  \int_{1}^{x}{\frac{W(u)}{u} du}}. 
\end{eqnarray}

\medskip

$\bullet$ The more accepted interpretation of the observational data is that the current universe is very close to a spatially flat geometry ($k = 0$), which seems to be a natural consequence from inflation in the early universe. In accordance with this, in the rest of the paper we will consider a flat 
 ($k = 0$) universe.

\medskip

 The Friedmann equation (\ref{field equations 1}) with $k = 0$ becomes 
\begin{equation}
\label{Hubble parameter, more general}
3H^2 = \frac{C}{x^3}\left[F(x) + G(x)\right].
\end{equation}
Thus, the density parameters are
\begin{equation}
\label{matter density parameter more general}
\Omega^{(M)} = \frac{ F(x)}{F(x) + G(x)}, \;\;\;\;\Omega^{(DE)} = \frac{ G(x)}{F(x) + G(x)}
\end{equation}
Since $F(1) = G(1) = 1$ it follows that   $\Omega^{(M)} = \Omega^{(DE)} = 1/2$ at  $x = 1$,  as expected. 
Now, taking the derivative we find 
\begin{equation}
\label{rate of Omega vs x}
\frac{d \Omega^{(M)}}{d x} = \frac{3 \left[W(x) - w(x)\right]}{x}\Omega^{(M)}\Omega^{(DE)}.
\end{equation}
This shows that $\Omega^{(M)}$ is a strictly decreasing function of $x$ as long as $\left[W(x) - w(x)\right] < 0$, which is always the case  for the mixture of DE $(W < 0)$ and  matter $(w \geq 0)$. It goes from some $1/2 < \Omega^{(M)}(0)  \leq 1$ to $\Omega^{(M)}(x \rightarrow \infty) \geq 0$. What this means is that  there is a one-to-one correspondence between $\Omega^{(M)}$ and
$x$. Therefore, if we know its value today  then (\ref{matter density parameter more general}) allows us to compute $x_{0}$ by solving the equation (as we mentioned before,  the subscript zero  denotes the quantities given  at the current epoch)
\begin{equation}
\label{x0 today, more general}
F(x_{0}) = \left[\frac{\Omega_{0}^{(M)}}{\Omega_{0}^{(DE)}}\right] G(x_{0}).
\end{equation}

\medskip

$\bullet$ The solution to this equation allows us to introduce the cosmological redshift $z$ in term of which $a = a_{0}/(1 + z)$. The relationship between $x$, defined in (\ref{definition of x}),  and $z$ is  
\begin{equation}
\label{x in terms of z}
z = \frac{x_{0}}{x} - 1.
\end{equation}
In particular, the transition from a matter-dominated era to a DE-dominated one occurs at the redshift $z_{*}$ given by
\begin{equation} 
\label{redshift of transition}
z_{*} = x_{0} - 1.
\end{equation}
Clearly, 
\begin{equation}
\label{x0 for different Omegas}
\left(\frac{a_{0}}{a_{*}}\right) = x_{0} \left\{\begin{array}{cc}
> 1, \;\;z_{*} > 0 & \mbox{for $\Omega_{0}^{(M)} < 1/2$}, \\
= 1, \;\;z_{*} = 0 & \mbox{for $\Omega_{0}^{(M)} = 1/2$}, \\
 < 1, \;\;z_{*} < 0 & \mbox{for $\Omega_{0}^{(M)} > 1/2$}.
\end{array}
\right.
\end{equation}
Thus, $x_{0} > 1$ indicates that the universe became DE dominated in the past;  $x_{0} < 1$ that DE will dominate in the future, and $x_{0} = 1$ that $\Omega_{0}^{(M)} =  \Omega_{0}^{(DE)} = 1/2$, i.e., that the DE-domination begins right now.

From (\ref{x0 today, more general}) and  (\ref{redshift of transition}) we find
\begin{equation}
\label{rate of Omega vs x, today today}
\frac{d x_{0}}{d\Omega_{0}^{(DE)}} = \frac{d z_{*}}{d\Omega_{0}^{(DE)}} = - \frac{x_{0}}{3\Omega_{0}^{(M)}\Omega_{0}^{(DE)}\left[W(x_{0}) - w(x_{0})\right]},
\end{equation}
which is always positive. 
This shows that the closer is $\Omega_{0}^{(DE)}$ to $1^{-}$, the higher  the redshift $z_{*}$ - i.e., the further back in time is   the transition to the DE-dominated era. 
 
\medskip 

Now we can evaluate the constant $C$. Using (\ref{Hubble parameter, more general}) and (\ref{x0 today, more general}) we find
\begin{equation}
C = \frac{3 H_{0}^2 x_{0}^3 \Omega_{0}^{(M)}}{F(x_{0})} = \frac{3 H_{0}^2 x_{0}^3 \Omega_{0}^{(DE)}}{G(x_{0})}.
\end{equation}
With this quantity at hand, the Hubble parameter (\ref{Hubble parameter, more general}) can be written as
\begin{equation}
\label{Hubble parameter, general}
H = H_{0}\left(\frac{x_{0}}{x}\right)^{3/2}\left[\Omega_{0}^{(M)} \frac{F(x)}{F(x_{0})} + \Omega_{0}^{(DE)} \frac{G(x)}{G(x_{0})}\right]^{1/2}.
\end{equation}

\medskip

$\bullet$ The deceleration parameter can be obtained from this expression by using 
\begin{equation}
q = - \frac{x}{H}\frac{d H}{d x} - 1,
\end{equation}
or directly from (\ref{general expression for q}) with $k = 0$. Either way we find
\begin{equation}
\label{aceleration in general}
q(x) = \frac{\left[1 + 3 w(x)\right] F(x) + \left[1 + 3 W(x)\right] G(x)}{2\left[F(x) + G(x)\right]}.
\end{equation}
In a similar manner, the deceleration parameter today is obtained immediately from (\ref{general expression for q}) with $k = 0$, as 
\begin{equation}
\label{q today, general}
q_{0} = \frac{1}{2}\left\{1 + 3\left[\Omega_{0}^{(M)} w(x_{0}) + \Omega_{0}^{(DE)} W(x_{0})\right]\right\}, 
\end{equation}
where $x_{0}$ is the solution to (\ref{x0 today, more general}) for a given $\Omega_{0}^{(M)}$.

The function $F(x)$ decreases,  and $G(x)$ increases, with the increase of $x$. Besides $W < - 1/3$. Therefore, at some point $q$ changes its sign from positive to negative. Let us use  $\bar{x}$ to denote the root(s) of the equation 
\begin{equation}
\label{equation for xbar}
q(\bar{x}) = \left[1 + 3 w(\bar{x})\right] F(\bar{x}) + \left[1 + 3 W(\bar{x})\right] G(\bar{x}) = 0.
\end{equation}
It should be noted that it may have several real roots because $q$ is not necessarily a monotonically decreasing function of $x$.   
In fact, from (\ref{aceleration in general}) we get
\begin{equation}
\label{derivative of q}
\frac{d q}{d x} = \frac{3 \Omega^{(M)}}{2}\frac{d w}{d x} - \frac{9 \Omega^{(M)}\Omega^{(DE)}}{2 x }\left[w - W\right]^2 + \frac{3 \Omega^{(DE)}}{2}\frac{d W}{d x}
\end{equation}
In this expression the first term is non positive, $\left(d w/d x\right) \leq 0$,  since the EoS of ordinary matter should become softer - or at most remain constant - during the expansion of the universe. The second one is negative. The third term could, in principle,  be positive, zero  or negative depending on the specific model. Therefore, an EoS for DE having a large positive slope may lead to several real solutions in (\ref{equation for xbar}) and the universe would pass from deceleration to acceleration, and vice versa, several times during its evolution.  In general, the only sure thing about the global behavior of $q$ is provided by the inequalities (\ref{limits on q, 1}) and (\ref{limits on q, 2}).
In summary, if $\left(d W/ d x\right) \leq 0$ then $q$ is a strictly decreasing function of $x$ and one can affirm  that (\ref{equation for xbar}) has only one solution.

Finally, we note that the acceleration of the universe at the end of the
  matter dominated phase, is given by\footnote{It is interesting to note that $q_{*}$ depends only on the specific values of the EoS at $x = 1$ and not on $F$ and $G$, which contain the  information on the overall effect of the EoS on the evolution of universe.} 
\begin{equation}
\label{q at the end of matter dominated era}
q_{*} = q(a_{*}) = q(x = 1) = \frac{1}{2} + \frac{3}{4}\left[w(1) + W(1)\right], 
\end{equation}
where we have used that $F(1) = G(1) = 1$. Clearly, if  $q_{*} < 0$  then the universe comes into an accelerated expansion, at some $\bar{x} < 1$, during   the period when   matter still dominates  over the dark energy component. In contrast, if  $q_{*} > 0$ then the accelerated expansion begins, at some $\bar{x} > 1$,  only after the DE component dominates over  matter. If  $q_{*} = 0$ then the onset of accelerated expansion coincides with the end of the matter-dominated era.

\section{Cosmological Model with variable EoS for matter and DE}

The formulae developed in the preceding sections give the cosmological parameters $\left(H, \Omega, q, r\right)$ explicitly in terms of $w$, $W$, $F$  
and $G$. Also (\ref{x0 today, more general})-(\ref{x in terms of z}) allow to express them as functions of the  redshift.  What we need  now, to construct a cosmological model, is to have the appropriate EoS.  In this section we do three things. First, we introduce  a variable EoS for the matter mixture (\ref{definition of rhom and pm}), and discuss its features. Second, we study a simple parameterization for DE
where $W$ varies between two asymptotic values. Third, we present the explicit expressions for the Hubble, density and deceleration parameters. 

\subsection{Effective EoS for the matter mixture}

In the approximation where relativistic matter is modeled by radiation and non-relativistic matter by dust, to describe the evolution of the universe from an early
radiation-dominated phase to the recent DE-dominated phase the function $w$ must satisfy the conditions  $w \rightarrow 1/3$  for  $x \ll 1$,  and $w \rightarrow 0$ for  $x \gg 1$.  
Certainly, there are an infinite number of smooth functions that satisfy these conditions.

To motivate our parameterization of $w$ let us go back to (\ref{definition of rhom and pm}).  The two main components of the cosmic mixture are provided by relativistic and non-relativistic particles. Therefore,
\[
w = \frac{\sum_{i} p^{(i)}}{\sum_{i} \rho^{(i)}} \approx \frac{p^{(NR)} + p^{(R)}}{\rho^{(NR)} + \rho^{(R)} },
\]
where $\rho^{(R)}$, $p^{(R)}$ and $\rho^{(NR)}$, $p^{(NR)}$ denote the energy density and pressure contributed by relativistic and non-relativistic particles, respectively. In the approximation under consideration $\rho^{(R)} = 3 p^{(R)}$ and $p^{(NR)} = 0$.  Thus, 
\[
w \approx \frac{1}{3 \left[f(x) + 1\right]}, \;\;\;f(x) = \frac{\rho^{(NR)}}{\rho^{(R)}}
\]
The relative contributions of the different components depend on time. During adiabatic expansion $\rho^{(NR)}$ decreases as $x^{- 3}$, while $\rho^{(R)}$ does it as $x^{- 4}$. Thus, $f(x) \sim x$ - at least - if one ignores interactions  where an individual kind of particle becomes non-relativistic, gets bound, or annihilates. In this paper we assume   $f(x) = K x ^{\alpha}$, where $K$ is a constant of proportionality and $\alpha $ is a positive constant parameter.  Thus,  we adopt  the following very
simple approximation 

\begin{equation}\label{3.1}
w = \frac{1}{3 \left(K x^{\alpha} + 1\right)}.
\end{equation}

\begin{figure}[tbp] 
  \centering
  \includegraphics[width=3.00in,height=3.00in,keepaspectratio]{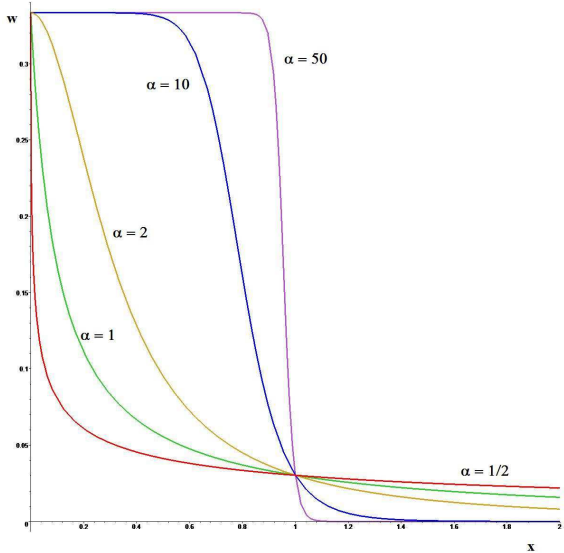}
  \caption{The effective EoS (\ref{3.1}) for the matter mixture (\ref{definition of rhom and pm}).  The Figure shows that for large values of $\alpha$ the transition from radiation to dust is very abrupt, and vice versa. 
The EoS at $x = 1$ is given by $w_{*} = 1/3\left(K + 1\right)$. For the sake of illustration, the graphic was constructed with $K = 10$.}
  \label{fig:EoSMatter}
\end{figure}

The  parameter $\alpha$ determines  the rapidity of the transition from radiation to dust as well as  the duration of the radiation-dominated epoch. In fact, differentiating with respect to $x$ we obtain
\begin{equation}
\label{rapidity of omega matter}
\frac{1}{w}\frac{d w}{d x} = - \frac{\alpha K \, x^{(\alpha - 1)}}{K x^{\alpha} +1}. 
\end{equation}
What this means is  that $w$ has a plateau, namely $w \approx 1/3$,  in the early universe  for any $\alpha > 1$. For $\alpha \leq 1$ there is a rapid variation of $w$ at $x \ll 1$ - there is no radiation-dominated plateau - and the transition to dust occurs very slowly. Besides, the larger the choice of $\alpha > 1$ the faster the  transition to the dust era $w \approx 0$, which goes on with the increase of $x$. See Fig. \ref{fig:EoSMatter}.

The coefficient $K$ determines the EoS at $x = 1$, viz.,
\[
w_{*} = w(x = 1) =  \frac{1}{3\left(K  + 1\right)}.
\]
To have $w_{*} \approx 0$ we assume that $K$ is a large, but finite, number. 

The mathematical simplicity of (\ref{3.1}) allows us to express all the relevant physical quantities  in terms of elementary functions. 
 In fact, substituting (\ref{3.1}) into (\ref{energy densities, general})  we find
\begin{equation}\label{3.3}
\rho^{(M)} = \frac{C_{1}\left(K x^{\alpha} +
1\right)^{1/\alpha}}{x^4}, \;\;\alpha \neq 0,\;\;\;C_{1} = \left(K + 1\right)^{- 1/\alpha} C.
\end{equation}
Consequently, 
\begin{equation}\label{3.4}
p^{(M)} = \frac{C_{1} \left(K x^{\alpha} + 1\right)^{\left(1 -
\alpha\right)/\alpha}}{3 x^4}.
\end{equation}

\medskip

$\bullet$ As expected,  for $\alpha = 1$ the constituents of the matter mixture recover their individual identities and $\rho^{(M)}$ separates into radiation and dust in adiabatic expansion, viz.,

\begin{equation}
\label{dust and radiation}
\rho^{(M)} = \frac{K C_{1}}{x^3} + \frac{C_{1}}{x^4}, \;\;\;\;p^{(M)} = \frac{C_{1}}{3 x^4}.
\end{equation}

\medskip

 For any $\alpha \neq 1$, the  EoS giving the relation between $p^{(M)}$ and $\rho^{(M)}$ is given in 
parametric form by (\ref{3.3}) and (\ref{3.4}). The explicit expression $p^{(M)} = p^{(M)}(\rho^{(M)})$ is quite cumbersome. However, the  asymptotic  behavior is as follows:

$\bullet$ For $x \ll 1$,
\[
\rho^{(M)} \approx  \frac{C_{1}}{x^4}, \;\;\;p^{(M)} \approx \frac{1}{3} \rho^{(M)},
\]

$\bullet$ For $x \gg 1$,

\[
\rho^{(M)} \approx  \frac{C_{1} \,K^{1/\alpha} }{x^3}, \;\;\;p^{(M)} = \frac{C_{1}\, K^{\left(1 - \alpha\right)/\alpha} }{3 x^{\left(3 + \alpha\right)}},
\]
For $\alpha > 0$, the matter pressure decreases with the expansion
of the Universe faster than the density.  Note that
$\rho^{(M)}$ decreases as in cosmological dust models. However, formally, {\it exact} dust models correspond to the limit $\alpha \rightarrow \infty$.  Bellow, in Section $5.1$,  we will discuss the properties of the models having  $\alpha \gg 1$.

$\bullet$
For $x \approx 1$, near the  transition density $\rho^{(M)}_{*}$ the EoS can be written as\footnote{From (\ref{3.3}) it follows that $y^4 - \left(C_{1}/\rho^{(M)}\right)^{\alpha} K y - \left(C_{1}/\rho^{(M)}\right)^{\alpha} = 0$ with $y \equiv x^{\alpha}$. At the transition poiny $x = 1$, $y = 1$, $\left(C_{1}/\rho^{(M)}_{*}\right)^{\alpha} = 1/\left(K + 1\right)$. In the neighborhood of $y = 1$ we can set $\left(C_{1}/\rho^{(M)}\right)^{\alpha} = 1/\left(K + 1\right) + \xi$ and $y = 1 + y_{1} \xi$, where $\xi$ is a dimensionless small parameter. Substituting into the equation for $y$ we get $y_{1} = \left(K + 1\right)^2/\left(3 K + 4\right)$, to first order in $\xi$.  Now $\xi = \left[\left(C_{1}/\rho^{(M)}\right)^{\alpha} - 1/\left(K + 1\right)\right]$ and using that $\left(C_{1}/\rho^{(M)}\right)^{\alpha} = \left(K + 1\right)^{- 1}\left(\rho_{*}^{(M)}/\rho^{(M)}\right)^{\alpha}$ we obtain $y = x^{\alpha} = \left(3 K + 4\right)^{- 1}\left[\left(2 K + 3\right) + \left(K + 1\right) \left(\rho_{*}^{(M)}/\rho^{(M)}\right)^{\alpha}\right]$. Finally, replacing this into (\ref{3.1}) and using (\ref{definition of omegas}) we get (\ref{chamgeover EoS}).
} 

\begin{equation}
\label{chamgeover EoS}
p^{(M)} \approx \frac{ \left(3 K + 4\right) \left[\rho^{(M)}\right]^{(\alpha + 1)}}{3 \left(K + 1\right)\left\{2\left(K + 2\right) \left[\rho^{(M)}\right]^{\alpha} + K \left[\rho^{(M)}_{*}\right]^{\alpha}\right\}}.
\end{equation}
This EoS closely mimics the behavior of the one discussed by Israelit and Rosen \cite{Israelit1}-\cite{Israelit2}   
where the pressure and density vary continuously  through different epochs in FRW universe models.

\subsection{A variable equation of state  for DE}

 To describe the evolution of DE we adopt the EoS   proposed by Hannestad and M\"{o}rtsell \cite{EoS}, which in our notation becomes\footnote{In our notation the four constants $w_{0}$, $w_{1}$, $q$ and $a_{s}$ in Hannestad-M\"{o}rtsell parameterization \cite{EoS} can be expressed as $q = \beta$, $w_{0} = {\omega}$, $w_{1} = \gamma$, $a_{s}^{\beta} = \left(w_{1}/w_{0}\right) a_{*}^{\beta}$.}
 
\begin{equation}\label{3.2}
W = \frac{{\omega} x^{\beta} + \gamma}{x^{\beta} + 1}, 
\end{equation}
where $\beta$, ${\omega}$ and $\gamma$ are constants parameters. 
If $\omega = \gamma$ then $W = \omega$. In any other case ${\omega}$ and $\gamma$ describe the asymptotic behavior of $W$. Namely, for $\beta > 0$  
\begin{equation}
\label{asymptotic behavior of omegade}
W \rightarrow \left\{\begin{array}{cc}
{\omega} & \mbox{for $x \gg 1$}, \\
\gamma & \mbox{for $x \ll 1$}.
\end{array}
\right.
\end{equation}
For $\beta < 0$, the constants ${\omega}$ and $\gamma$ interchange their role, i.e., $W \rightarrow \gamma$ for $x \gg 1$ 
and $W \rightarrow {\omega}$  for $x \ll 1$. In what follows without loss of generality we assume $\beta > 0$. 

 Besides,  
\[
\frac{d W}{d x} = \frac{\beta x^{\beta - 1} \left(\omega - \gamma\right)}{\left(x^{\beta} + 1\right)^2}
\]
shows that for $\beta > 1$,  $W$ undergoes a rapid transition from an early epoch 
where $W \approx \gamma$ to a large epoch where $W \approx {\omega}$. In addition, $W$ is an increasing function of $x$ if ${\omega} > \gamma$, and 
vice versa.

Next, to account for the repulsive nature of DE we should demand the violation of the strong energy condition of classical cosmology. It is easy to verify that   $\rho^{(DE)} + 3 p^{(DE)} = \rho^{(DE)}(1 + 3 W) < 0$ in the whole range of $x$ provided 
\begin{equation} 
\label{strong energy condition for DE}
{\omega} < - \frac{1}{3}, \;\;\;\gamma < - \frac{1}{3}.
\end{equation}
Lower 
limits on these quantities, namely ${\omega} > - 1$ and $\gamma > - 1$ are obtained if one assumes that DE satisfies the dominant energy condition $\rho^{(DE)} \geq |p^{DE}|$.

Substituting (\ref{3.2}) into (\ref{energy densities, general}), we obtain\footnote{The choice $\beta = 0$ correspond to a constant equation of state $W = \left({\omega} + \gamma\right)/2$.}
\begin{equation}\label{3.5}
\rho^{(DE)} = \frac{C_{2}\left(x^{\beta} + 1\right)^{-
3({\omega} - \gamma)/\beta}}{x^{3 (1 + \gamma)}},\;\; \beta \neq 0,\;\;\;C_{2} = 2^{3\left(\omega - \gamma\right)/\beta}\, C.
\end{equation}
Thus,
\begin{equation}\label{3.6}
p^{(DE)} = \frac{C_{2} \left({\omega} x^{\beta} + \gamma\right)}{x^{3\left(1 + \gamma\right)} \left(x^{\beta} + 1\right)^{[3 \left({\omega} - \gamma\right) + \beta]/\beta}}.
\end{equation}

We immediately notice that for $\beta = 3\left(\gamma - \omega\right)$ the above equations reduce to 
\begin{eqnarray}
\label{density and pressure, superposition of fluids}
\rho^{(DE)} &=& C_{2}\left[\frac{1}{x^{3 \left(1 + \omega\right)}} + \frac{1}{x^{3\left(1 + \gamma\right)}}\right],\nonumber \\
p^{(DE)} &=& C_{2}\left[\frac{\omega}{x^{3 \left(1 + \omega\right)}} + \frac{\gamma}{x^{3\left(1 + \gamma\right)}}\right].
\end{eqnarray}
What this means is that for $\beta = 3\left(\gamma - \omega\right)$ the DE can be  interpreted as the superposition of two fluids with EoS 
\begin{equation}
p_{1} = \omega \rho_{1}, \;\;\;\;p_{2} = \gamma \rho_{2},
\end{equation}
each of which satisfies the continuity equation.

\medskip

 The asymptotic behavior of (\ref{3.5})-(\ref{3.6}) is as follows: 

\medskip

$\bullet$ For $x \ll 1$, there are two different physical situations. If $\gamma \neq 0$ we get
\[
 \rho^{(DE)} \approx \frac{C_{2}}{x^{3 (1 + \gamma)}}, \;\;\;p^{(DE)} \approx \gamma \rho^{(DE)}.
\]
In this limit $\rho^{(DE)}/\rho^{(M)} \approx x^{1 - 3 \gamma}$.  Therefore $\rho^{(DE)} \ll \rho^{(M)} $ for $x \ll 1$ as long as $\gamma < 1/3$. 

If  $\gamma = 0$ we find 
\[
 \rho^{(DE)} \approx \frac{C_{2}}{x^{3 }}, \;\;\;p^{(DE)} \approx \omega C_{2} x^{\left(\beta - 3\right)}.
\]
Thus, for $\beta > 3$ the DE component - in this limit -  behaves exactly as in cosmological dust models. 

\medskip

$\bullet$ For $x \gg 1$ we obtain
\[
\rho^{(DE)} \approx \frac{C_{2}}{x^{3\left(1
+ {\omega}\right)}}, \;\;\;p^{(DE)} = {\omega} \rho^{(DE)}.
\]
Since ${\omega} <  0$, it follows that $\rho^{(DE)} \gg \rho^{(M)}$ and
$|p^{(DE)}| \gg p^{(M)}$ for $x \gg 1$. Also, in this limit when ${\omega} = -1$ the DE component  behaves as a cosmological constant $p^{(DE)} = - \rho^{(DE)} =  C_{2}$. 

\medskip 

$\bullet$ For $x \approx 1$, the transition 
from $W \approx \gamma$ to  $W  \approx {\omega}$ can be  approximated by an EoS similar to  (\ref{chamgeover EoS}).
\paragraph{Generalized Chaplygin gas:} 

The above analysis indicates that when ${\omega} = -1$, $\beta > 3$ and $\gamma = 0$ the DE component evolves from non-relativistic matter (dust) to a cosmological constant, similar to the so-called Chaplygin gas\footnote{Chaplygin gas satisfies the EoS equation of state $ p = - A/\rho^{\sigma}$ where $p$ is the pressure, $\rho$ is the density, with $\sigma = 1$ and $A$ a positive constant.  In generalized Chaplygin gas models $\sigma$ is a parameter which can take on values $0 < \sigma < 1$}.  In fact, it is not difficult to see that for this choice of parameters (\ref{3.5})-(\ref{3.6}) yield 
\begin{equation}
\label{EoS for Chaplygin gas}
p^{(DE)}\left[\rho^{(DE)}\right]^{\sigma} = - C_{2}^{\left(1 + \sigma\right)},
\end{equation}
where $\sigma = (\beta - 3)/3$ and 
\begin{equation}
\label{Chaplygin gas}
\rho^{(DE)} = C_{2}\left[1 + \frac{1}{x^{3 \left(1 + \sigma\right)}}\right]^{1/\left(1 + \sigma\right)}.
\end{equation}
Thus, the EoS 
\begin{equation}
W = - \frac{x^{\beta}}{x^{\beta} + 1}, 
\end{equation}
with $3 < \beta < 6$ describes a generalized Chapligyn gas.

\subsection{The Hubble, density and deceleration parameters }

We now proceed to apply our general formulae to the EoS (\ref{3.1}) and (\ref{3.2}). For practical reasons, it is useful to write (\ref{3.1}) as 
\begin{equation}
\label{EoS with n}
w = \frac{n}{ \left(K x^{\alpha} + 1\right)},
\end{equation}
with n = 1/3. This will allow us to follow in detail, step by step,  the possible effects of the primordial radiation on the observed accelerated expansion today. Besides, it allows us to extend our results to  include other types of matter, e.g., stiff (incompressible) matter ($n = 1$). Also for $n = 0$ we should recover the usual CDM picture. 

Thus, for the sake of generality in our discussion we let 
\begin{equation}
0 \leq n \leq 1,
\end{equation} 
in accordance with the strong and dominant energy conditions. 

From (\ref{definition of F and G}) we get
\begin{equation}
\label{explicit forms of F and G}
 F(x) = \left(K + 1\right)^{- 3 n/\alpha}\left[K + \frac{1}{x^{\alpha}}\right]^{3 n/\alpha}, \;\;\;G(x) =  \frac{2^{3\left(\omega - \gamma\right)/\beta}}{x^{3 \omega}} \left[1 + \frac{1}{x^{\beta}}\right]^{- 3\left(\omega - \gamma\right)/\beta}.
\end{equation}

\medskip

$\bullet$ The Hubble parameter is obtained by substituting these  into (\ref{Hubble parameter, general}), namely,  
\begin{equation}
\label{Hubble parameter for our problem}
H(x) = H_{0}\left[\Omega_{0}^{(M)}\left(\frac{x_{0}}{x}\right)^3\, f(x) + \Omega_{0}^{(DE)} \left(\frac{x_{0}}{x}\right)^{3\left(\omega + 1\right)} g(x)\,\right]^{1/2},
\end{equation}
where we have introduced the functions $f(x)$ and $g(x)$ defined by 
\begin{equation}
\label{f and g for our problem}
f(x) = \left[\frac{K + \frac{1}{x^{\alpha}}}{K + \frac{1}{x^{\alpha}_{0}}}\right]^{3 n/\alpha}, \;\;\;\;g(x) = \left[\frac{1 + \frac{1}{x_{0}^{\beta}}}{1 + \frac{1}{x^{\beta}}}\right]^{3 \left(\omega - \gamma\right)/\beta}, 
\end{equation}
and $x_{0}$ is the solution to (\ref{x0 today, more general}), which in the case under consideration becomes
\begin{equation}
\label{definition of x0}
x_{0}^{3 {\omega}}\left(K + \frac{1}{x_{0}^{\alpha}}\right)^{3 n/\alpha}
\left(1 + \frac{1}{x_{0}^{\beta}}\right)^{3({\omega} - \gamma)/\beta} = 2^{3({\omega} - \gamma)/\beta} \left(K + 1\right)^{3 n/\alpha}\left[\frac{\Omega_{0}^{(M)}}{\Omega_{0}^{(DE)}}\right].
\end{equation} 

The energy densities, from (\ref{energy densities, general}), (\ref{definition of F and G}), (\ref{explicit forms of F and G}), (\ref{f and g for our problem}),  can be written as 
\begin{eqnarray}
\label{final form for the densities}
\rho^{(M)} &=& \rho^{(M)}_{0}\left(\frac{x_{0}}{x}\right)^{3} f(x),\nonumber \\
\rho^{(DE)} &=& \rho^{(DE)}_{0}\left(\frac{x_{0}}{x}\right)^{3\left(1 + \omega\right)} g(x).
\end{eqnarray}
For $n = 0$ and $\gamma = \omega$, these expressions reduce to the usual ones in the CDM model,  as expected.

\medskip

$\bullet$ The deceleration parameter is easily obtained by a direct substitution of (\ref{3.2}), (\ref{EoS with n}) and (\ref{explicit forms of F and G}) into (\ref{aceleration in general}). We get
\begin{equation}
\label{general expression for q in our problem}
q(x) = \frac{x^{3 \omega}\left(K + \frac{1}{x^{\alpha}}\right)^{3 n/\alpha}   \left[\frac{K x^{\alpha} + 1 + 3 n}{K x^{\alpha} + 1}\right]   + 2^{3\left(\omega - \gamma\right)/\beta}\left(K + 1\right)^{3 n/\alpha} \left(1 + \frac{1}{x^{\beta}}\right)^{- 3\left(\omega - \gamma\right)/\beta}\left[\frac{\left(3 \omega + 1\right) x^{\beta} + \left(3 \gamma + 1\right)}{x^{\beta} + 1}\right]}{2 \left[x^{3 \omega}\left(K + \frac{1}{x^{\alpha}}\right)^{3 n/\alpha} + 2^{3\left(\omega - \gamma\right)/\beta }\left(K + 1\right)^{3 n/\alpha}\left(1 + \frac{1}{x^{\beta}}\right)^{- 3\left(\omega - \gamma\right)/\beta}\right]}.
\end{equation}

\begin{figure}[tbp] 
  \centering
  \includegraphics[width=3.00in,height=3.00in,keepaspectratio]{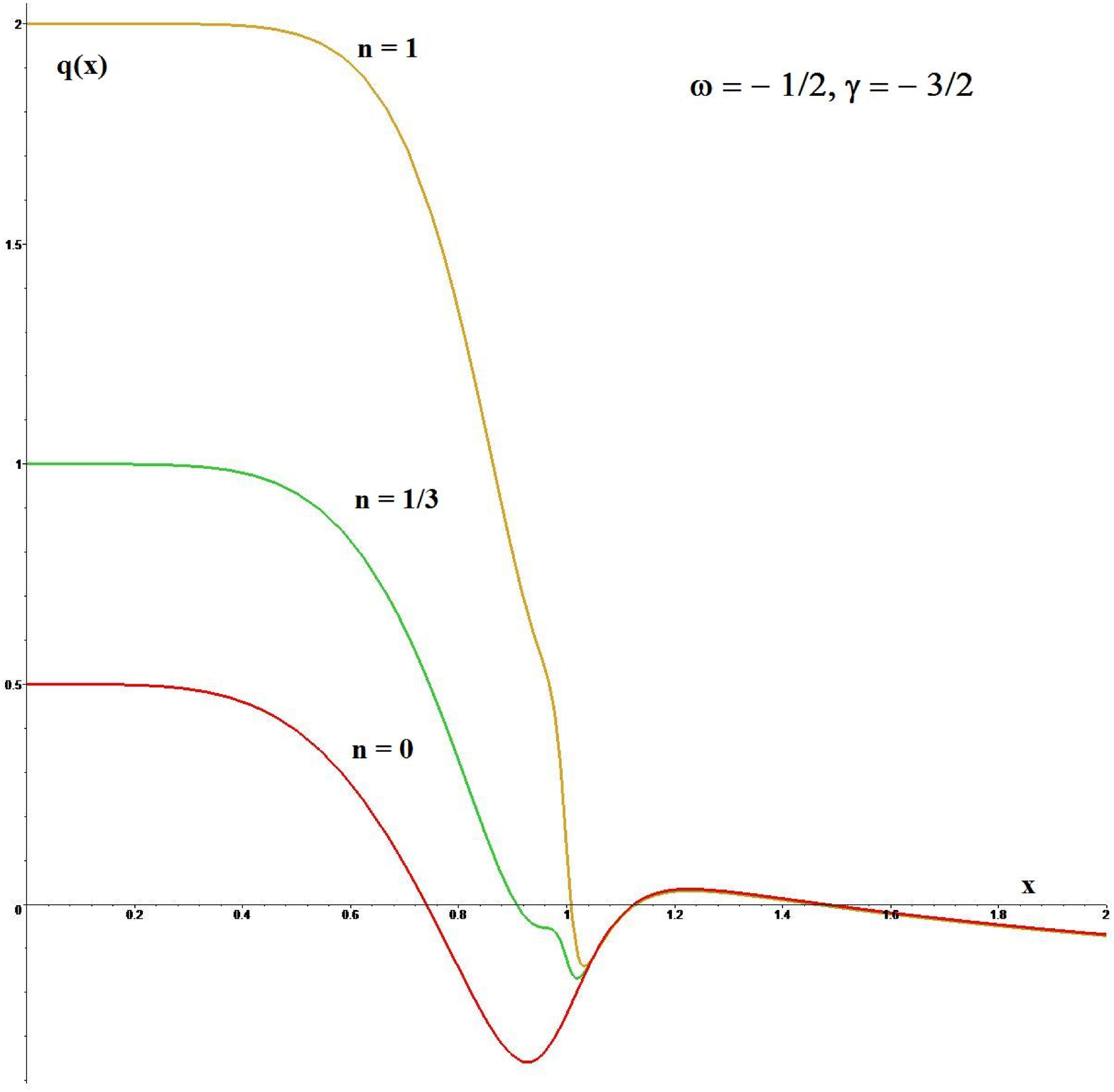}
  \caption{The figure illustrates the situation where the deceleration parameter changes sign several times during the expansion of the universe. We have selected $\omega = - 1/2$, $\gamma = - 3/2$ and $\beta = 20$ to generate a large $d W/ d x > 0$.  Also, we have chosen $\alpha = 100$ to assure that the models with primordial radiation ($n = 1/3$) and stiff matter $(n = 1)$ are indistinguishable from the dust model at late times. Also, for this choice the number of zeroes of $q(x)$ as well as their locations are essentially independent of the specific value of   $K$. 
The graphic was constructed with $K = 1$. }
  \label{fig:q0}
\end{figure}

We observe that  $q(x = 0) = \left(1 + 3 n\right)/2$ and $q(x \rightarrow \infty) = \left(1 + 3 {\omega}\right)/2$.  Since $\omega < - 1/3$, there is at least one value of $x$, say $x = \bar{x}$,  for which $q$ vanishes\footnote{In Section $5.3$ we will consider models with $\omega = 0$.}. In accordance with our discussion in (\ref{derivative of q}),  $q$ is  not necessarily a decreasing function of $x$ if $d W/d x > 0$ in which case $q(\bar{x}) = 0$ may have more than one real root. In Figure \ref{fig:q0} we illustrate the  situation where $q(x) = 0$ has three different roots for different EoS: dust ($n = 0$), primordial radiation ($n = 1/3$) and stiff matter ($n = 1$).

In general, only the upper and lower bounds on $q$ are well-defined and are given by (\ref{limits on q, 1}). In the case under consideration these are  
\begin{equation}
\frac{1 + 3 \xi}{2} \leq q \leq 1, \;\;\;\xi = \mbox{min}\left\{\gamma, {\omega}\right\}. 
\end{equation}
 Besides,  in agreement with (\ref{q at the end of matter dominated era}) we find that at the transition point $x = 1$
 ($a = a_{*}$)  
\begin{equation}
q_{*} \equiv q(x = 1) = \frac{1}{2} + \frac{3\left(\omega + \gamma\right)}{8} + \frac{3 n}{4 \left(K + 1\right)},
\end{equation}
  regardless of the specific choice  of $\alpha$ and $\beta$.

\medskip

$\bullet$ The magnitude of the deceleration parameter today, $q_{0}$, is readily given by (\ref{q today, general}), viz., 

\begin{equation}
\label{q today}
q_{0} = \frac{1}{2}\left[1 + \frac{3 n \Omega_{0}^{(M)}}{K x_{0}^{\alpha} + 1} + \frac{3 \Omega_{0}^{(DE)}\left({\omega} x_{0}^{\beta}+ \gamma\right)}{x_{0}^{\beta} + 1}\right].
\end{equation}
The same expression can be obtained from (\ref{general expression for q in our problem}), after using (\ref{definition of x0}) and some algebraic manipulations.

\subsubsection{Cosmological parameters in terms of the redshift}

For practical/observational  reasons it is convenient to express  the cosmological parameters  as functions of the redshift $z$  introduced in (\ref{x in terms of z}).   

The Hubble parameter (\ref{Hubble parameter for our problem}) becomes

\begin{equation}
\label{H in terms of z}
H^2 = H_{0}^2 \left[\Omega_{0}^{(M)}\left(1 + z\right)^3 f(z) + \Omega_{0}^{(DE)}\left(1  + z\right)^{3(\omega + 1)} g(z)\right], 
\end{equation}
with
\begin{eqnarray}
\label{definition of f and g in terms of z}
f(z) &=& \left[\frac{K x_{0}^{\alpha} + \left(1 + z\right)^{\alpha}}{K x_{0}^{\alpha} + 1}\right]^{3 n/\alpha}, \nonumber \\
g(z) &=& \left[\frac{x_{0}^{\beta} + \left(1 + z\right)^{\beta}}{x_{0}^{\beta} + 1}\right]^{- 3 \left({\omega} - \gamma\right)/\beta}.
\end{eqnarray}

The energy densities can be written as 
\begin{eqnarray}
\rho^{(M)} &=& 3\left(1 + z\right)^3 \Omega_{0}^{(M)} H_{0}^2 f(z), \nonumber \\
\rho^{(DE)} &=& 3 \left(1 + z\right)^{3\left(1 + {\omega}\right)}\Omega_{0}^{(DE)} H_{0}^2 g(z).
\end{eqnarray}

Finally,  the deceleration parameter as a function of $z$ is given by

\begin{equation}
\label{q general in terms of z}
q(z) = \frac{H_{0}^2}{2 H^2}\left\{\Omega_{0}^{(M)}\left(1 + z\right)^3 f(z)\frac{\left[K x_{0}^{\alpha} + \left(3 n + 1\right)\left(1 + z\right)^{\alpha}\right]}{\left[K x_{0}^{\alpha} + \left(1 + z\right)^{\alpha}\right]} + \Omega_{0}^{(DE)}\left(1 + z\right)^{3\left(1 + {\omega}\right)} g(z)\frac{\left[x_{0}^{\beta}\left(1 + 3{\omega}\right) + \left(1 + z\right)^{\beta} \left(1 + 3 \gamma\right)\right]}{\left[x_{0}^{\beta} + \left(1 + z\right)^{\beta}\right]}\right\}.
\end{equation}
These equations show that at low redshift $f(z) \approx 1$ and $g(z) \approx 1$ and we recover the usual CDM model, i.e., the dynamics of the  EoS for DE  and matter  only becomes important at hight redshift.

\section{Universe of matter with variable EoS and DE with $W = $ constant}

We now start the study of the properties of our cosmological model. In this section we concentrate our attention  on the question of how a dynamical EoS for matter can alter the details  of the accelerated expansion as given by the CDM ($w = 0$) models. With this in mind, here we confine the discussion to the case where  the DE component has a constant EoS (a variable $W$ will be discussed in the next section).

Thus, in this section    
\begin{equation}
\label{W constant}
W = \omega = \mbox{constant}.
\end{equation} 
For $\omega = - 1$ we have a standard cosmological constant.
Setting $\gamma = \omega$, from (\ref{definition of x0}), (\ref{general expression for q in our problem}) and (\ref{q today}) we get 
\begin{equation}
\label{x0 with n}
x_{0}^{3\omega}\left(K + \frac{1}{x^{\alpha}_{0}}\right)^{3 n/\alpha} = \left(K + 1\right)^{3 n/\alpha}\frac{\Omega_{0}^{(M)}}{\Omega_{0}^{(DE)}}, 
\end{equation}

\begin{equation}
\label{eq. for x bar, case with n}
q(\bar{x}) = \bar{x}^{3\omega}\left[K + \frac{1}{\bar{x}^{\alpha}}\right]^{3 n/\alpha}\left[\frac{K \bar{x}^{\alpha} + 1 + 3 n}{K \bar{x}^{\alpha} + 1}\right] + \left(K + 1\right)^{3 n/\alpha}\left(1 + 3\omega\right) = 0, 
\end{equation}
and 

\begin{equation}
\label{q0 for the model with n}
q_{0} = \frac{1}{2}\left[1 + 3\omega \Omega_{0}^{(DE)}\right] + \frac{3 n \Omega_{0}^{(M)}}{2\left(K x_{0}^{\alpha} + 1\right)}, 
\end{equation}
respectively. Thus, to obtain a numerical value for $q_{0}$ as well as the observational quantities $z_{*} = x_{0} - 1$ and $\bar{z} = x_{0}/\bar{x} - 1$, we need to solve (\ref{x0 with n}) and (\ref{eq. for x bar, case with n}).  

\paragraph{CDM:} For $n = 0$, they have a close algebraic solution which is the CDM model, viz.,
\begin{eqnarray}
\label{CDM solution}
x_{0}^{(CDM)} &=& \left[\frac{\Omega_{0}^{(M)}}{\Omega_{0}^{(DE)}}\right]^{1/3\omega}, \nonumber \\
{\bar{x}}^{(CDM)} &=& \left[ - \left(1 + 3 \omega\right)\right]^{1/3\omega}, \nonumber \\
q_{0}^{(CDM)} &=& \frac{1}{2}\left[1 + 3\omega \Omega_{0}^{(DE)}\right]. 
\end{eqnarray}

\paragraph{Variable EoS:} For $n \neq 0$,  from (\ref{x0 with n})-(\ref{eq. for x bar, case with n}) we find expressions for $x_{0}$ and $\bar{x}$ which are very similar to those given by (\ref{CDM solution}), namely
 
\begin{eqnarray}
x_{0} &=& \left(\frac{w_{0}}{w_{*}}\right)^{n/\alpha\left(\omega - n\right)}\left[\frac{\Omega_{0}^{(M)}}{\Omega_{0}^{(DE)}}\right]^{1/3\left(\omega - n\right)}, \nonumber \\
\bar{x} &=& \left(\frac{\bar{w}}{w_{*}}\right)^{n/\alpha\left(\omega - n\right)}\left[- \frac{1 + 3 \omega}{1 + 3 \bar{w}}\right]^{1/3\left(\omega - n\right)},
\end{eqnarray}
where $w_{0}$, $w_{*}$ and $\bar{w}$ represent the EoS (\ref{EoS with n}) evaluated at $x_{0}$, $x = 1$ and $x = \bar{x}$, respectively. They reduce to their CDM counterpart for  $n = 0$, as expected. Although the shape of these expressions is intriguing,  
to find  $x_{0}$ and $\bar{x}$ we need to provide a set of values $(\Omega_{0}^{(M)},\, \omega, \,n,\, K, \, \alpha)$ and then solve numerically.

However, a direct  inspection of (\ref{x0 with n})-(\ref{eq. for x bar, case with n}) suggests that we examine the following cases: (i) $\alpha = 3n$; (ii) $0 <\alpha \ll 3 n$ and (iii) $\alpha \gg 3 n$.

\medskip 

$\bullet$ When $\alpha = 3 n$, the matter in the universe is a superposition of two non-interacting fluids; one of them is dust, and the other one is a fluid with an EoS $p = n \rho$. Specifically,  
\begin{equation}
\rho^{(M)} = \frac{K C_{1}}{x^3} + \frac{C_{1}}{x^{3\left(n + 1\right)}}, \;\;\;\;p^{(M)} = \frac{n C_{1}}{x^{3\left(n + 1\right)}}.
\end{equation}
 
\medskip

$\bullet$ When $0 < \alpha \ll 3n$ we have
\[
\rho^{(M)} = \frac{C_{1}}{x^{3\left(1 + \tilde{n}\right)}},\;\;\;\; \tilde{n} = \frac{n}{K + 1},
\]
\begin{equation}
x_{0} = \left[\frac{\Omega_{0}^{(M)}}{\Omega_{0}^{(DE)}}\right]^{1/3\left(\omega - \tilde{n}\right)},
\end{equation}
\begin{equation}
\bar{x} = \left[\frac{- \left(1 + 3 \omega\right)}{1 + 3 \tilde{n}}\right]^{1/3\left(\omega - \tilde{n}\right)}.
\end{equation}
Since $K$ is a large number ($K \sim \rho^{(NR)}/\rho^{(R)} \sim 3 \times 10^4$ today), the above solution is practically indistinguishable from the CDM model (\ref{CDM solution}).

$\bullet$ When $\alpha \gg 3 n$,  we find that the physics is independent 
of the particular choice of $K$ and $\alpha$. In the next subsection we discuss the details.

\subsection{Asymptotic model: $\alpha \gg 3n$}

A simple numerical analysis of (\ref{x0 with n})-(\ref{eq. for x bar, case with n}),  with  $(\Omega_{0}^{(M)},\, \omega, \,n,\, K)$ fixed,  shows that $x_{0}$ and $\bar{x}$ tend to  specific finite limits  when $\alpha \rightarrow \infty$. A further analysis demonstrates that these limits are insensitive\footnote{In fact, one can show that $\frac{1}{x_{0}}\frac{d x_{0}}{d K} \sim \frac{1}{\bar{x}}\frac{d \bar{x}}{d K} \sim \frac{n}{\alpha \left(K + 1\right)}$ so that $\frac{1}{x_{0}}\frac{d x_{0}}{d K} \sim \frac{1}{\bar{x}}\frac{d \bar{x}}{d K} \rightarrow 0$ as $\alpha \rightarrow \infty$.} to the specific choice of $K$.

\begin{figure}[tbp] 
  \centering
  \includegraphics[width=3.00in,height=3.00in,keepaspectratio]{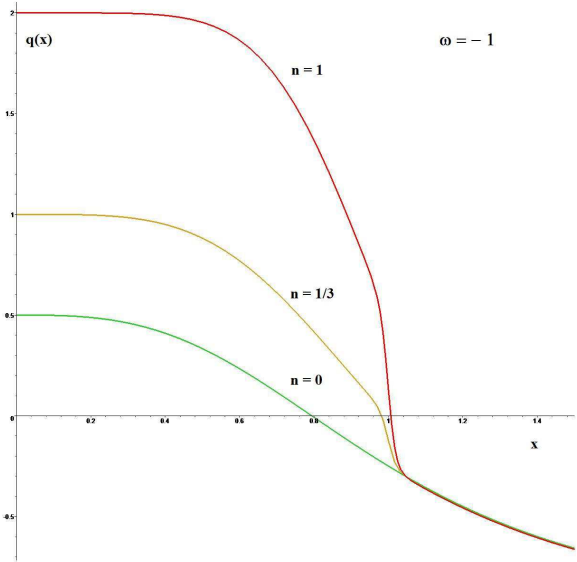}
  \caption{The deceleration parameter  $q$ versus $x$ for models with cosmological constant. The figure compares the evolution of $q$ in the $\Lambda$CDM model ($n = 0$) to the one in models with $n = 1/3$ and  $n = 1$ calculated for $\alpha = 100$ and $K = 1$.}
  \label{fig:q1}
\end{figure}

\begin{figure}[tbp] 
  \centering
  \includegraphics[width=3.00in,height=3.00in,keepaspectratio]{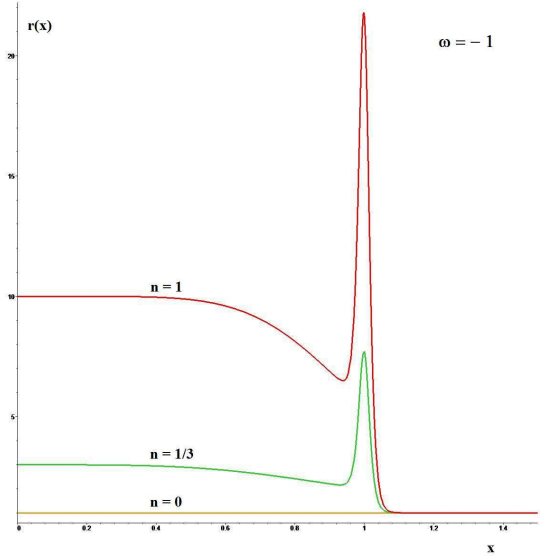}
  \caption{The statefinder parameter $r$, given by (\ref{general expression for r}),  versus $x$ for models with cosmological constant. In the $\Lambda$CDM models $r = 1$. The sharp peak of $r$ is evidence of the rapid change experienced by the deceleration parameter. Once again we have taken $\alpha = 100$ to ensure that the models behave like dust at late times. Moreover $K = 1$.}
  \label{fig:r1}
\end{figure}

Therefore, in this limit only the parameters $\omega$, $n$ and $\Omega_{0}^{(M)}$ have physical relevance. It corresponds to the case where the transition to dust occurs abruptly  near $x = 1$ (See Figures \ref{fig:EoSMatter}, \ref{fig:q1}, \ref{fig:r1} ). 

As shown in (\ref{x0 for different Omegas}), the character of the solutions to (\ref{x0 with n}) strongly depends on whether the universe is dominated by matter or DE.  Accordingly,  for $\alpha \gg 1$ we find

\begin{equation}
\label{solution for x0 and different Omegas, 1}
 x_{0} =  \left\{\begin{array}{cc}
  x_{0}^{(CDM)}& \mbox{for $\Omega_{0}^{(M)} < 1/2$}, \\
\left[\frac{\Omega_{0}^{(M)}}{\Omega_{0}^{(DE)}}\right]^{1/3\left(\omega - n\right)} & \mbox{for $\Omega_{0}^{(M)} > 1/2$}. 
 \end{array}
\right.
\end{equation}
For $\Omega_{0}^{(M)} = 1/2$, $x_{0} = 1$ regardless of the choice of parameters.
To simplify the discussion in Table $1$ we calculate $x_{0}$ for various $\Omega_{0}^{(0)} < 1/2$ and $\omega$. 

\begin{center}
\begin{tabular}{|c|c|c|c|c|c|c|c|} \hline
\multicolumn{8}{|c|}{\bf Table 1. $x_{0}$ for various $\Omega_{0}^{(M)}$, $\omega$, any $n$ and $\alpha \gg 1$}\\ \hline
 $\Omega_{0}^{(M)}$   & $ \omega = - 0.4$& $ \omega = - 0.5$& $\omega = - 2/3$& $ \omega = - 1$ &$\omega = - 1.3$& $\omega = - 1.5$&$\omega = - 10$    \\ \hline
$ 0.1$ & $6.240$& $4.327$& $3.000$& $2.080$ &$1.756$ &$1.629$&$1.076$   \\ \hline
$ 0.2$ & $3.175$& $2.520$& $2.000$& $1.587$ &$1.427$ &$1.361$&$1.047$   \\ \hline
$ 0.3$ & $2.026$& $1.759$& $1.527$& $1.326$ &$1.243$ &$1.207$&$1.028$   \\ \hline
$ 0.4$ & $1.402$& $1.310$& $1.225$& $1.148$ &$1.110$ &$1.094$&$1.014$   \\ \hline
\end{tabular}
\end{center}

\medskip

$\bullet$  By virtue of (\ref{derivative of q}), the EoS (\ref{W constant}) guarantees that $q(x)$ is a strictly decreasing function of $x$. Consequently, (\ref{eq. for x bar, case with n}) has only one real root  for each set $(\omega, \, n,\, K, \alpha)$. 
They heavily depend on whether the deceleration-acceleration  transition occurs before or after the end of the matter-dominated era. From  (\ref{q at the end of matter dominated era}) we find
\begin{equation}
q_{*} = \frac{1}{4}\left(2 + 3 \omega + \frac{3 n}{K + 1}\right).
\end{equation}
What this means is that $\bar{x} \geq 1$ for $\omega \geq - \left[2/3 + n/(K + 1)\right]$,  and vice versa.
 
Thus, for $\alpha \gg 1$ we obtain the following solutions: 

\paragraph{(i) $\omega > - 2/3$:}

\[
\bar{x} = \left[ - \left(1 + 3 \omega\right)\right]^{1/3\omega},
\]
where the condition $\omega > - 2/3$ ensures  that $\bar{x} > 1$.

\paragraph{(ii) $ - \left[2/3 + n/(K + 1)\right] < \omega \leq - 2/3$:}
\[
\bar{x} = 1^{+}.
\]
\paragraph{(iii) $\omega =  - \left[2/3 + n/(K + 1)\right]$:}
\[
\bar{x} = 1,
\]
for any choice of parameters.
\paragraph{(iv) $- \left(2 + 3 n\right)/ 3 \leq \omega < - \left[2/3 + n/(K + 1)\right]$:}
\[
\bar{x} = 1^{-}.
\]

\paragraph{(v) $\omega < - \left(2 + 3 n\right)/3$:}

\[
\bar{x} = \left[- \frac{\left(1 + 3 \omega\right)}{\left(1 + 3 n\right)}\right]^{1/3\left(\omega - n\right)}, 
\]
where the condition $\omega < - \left(2 + 3 n\right)/3$ ensures that $\bar{x} < 1$.

For $n = 0$ we recover the expressions for CDM. In Table $2$ we present the approximate values of $\bar{x}$ for different choices of $\omega$, $n$ and $ \alpha \gg 1$.  We remark that those for $\omega \geq - 2/3$ are independent of $n$ and coincide with the CDM values.

\bigskip

\begin{center}
\begin{tabular}{|c|c|c|c|c|c|c|} \hline
\multicolumn{7}{|c|}{\bf Table 2. $\bar{x}$ for various $\omega$, $n$ and $\alpha \gg 1$}\\ \hline
 $ \omega$ & $ n = 0$& $ n = 0.1$& $n = 0.2$& $ n = 1/3$ &$n = 2/3$& $n = 1$    \\ \hline
$ - 0.4$ & $3.824$& $3.824$& $3.824$& $3.824$ &$3.824$ &$3.824$   \\ \hline
$ - 0.5$ & $1.587$& $1.587$& $1.587$& $1.587$ &$1.587$ &$1.587$   \\ \hline
$ - 2/3$ & $1$& $1$& $1$& $1$ &$1$ &$1$   \\ \hline
$ - 5/6$ & $0.850$& $0.950$& $ 1$& $1$ &$1$ &$1$   \\ \hline
$ - 1$ & $0.794$& $0.878$& $0.940$& $1$ &$1$ &$1$   \\ \hline
$ - 1.3$ & $0.761$& $0.826$& $0.876$& $0.927$ &$1$ &$1$   \\ \hline
$ - 1.5$ & $0.757$& $0.813$& $0.858$& $0.903$ &$0.976$ &$1$   \\ \hline
$ - 10$ & $0.894$& $0.903$& $0.910$& $0.917$ &$0.932$ &$0.942$   \\ \hline
\end{tabular}
\end{center}

\bigskip

If $\omega < - 2/3$, then our model predicts some changes to the CDM paradigm. First we note that  in CDM models (first column in Table $2$)  $\bar{x}$ varies with $\omega$. But, this is not necessarily so when $n \neq 0$. Solutions (ii)-(iv)  show  that $\bar{x} \approx 1$, for every $n$,  in the whole range  $- \left(2 + 3 n\right)/3 \leq \omega \leq - 2/3$.  

However, the redshift associated with $\bar{x}$ does depend on $\omega$ through $x_{0}$. For example, if we take $n = 1/3$ and $\Omega_{0}^{(M)} = 0.3$ then Table $1$ gives $x_{0} = \left(1.527, \; 1.326\right)$ for $\omega = - 2/3$ and $\omega = - 1$, respectively. Correspondingly,  we get $\bar{z} = \left(0.527, \; 0.326\right)$, although $\bar{x} = 1$ in both cases. 

When $\omega < - \left(2 + 3 n\right)/ 3$, for $\Omega_{0}^{(M)} < 1/2$ the redshift for the deceleration-acceleration 
transition is given by\footnote{For $\Omega_{0}^{(M)} > 1/2$ we should use the second solution in (\ref{solution for x0 and different Omegas, 1}). See also (\ref{q = 0 transition for matter dominated universe}).}  
\begin{equation}
\bar{z} = \left[\frac{\Omega_{0}^{(M)}}{\Omega_{0}^{(DE)}}\right]^{1/3\omega}\left[- \frac{\left(1 + 3 \omega\right)}{\left(1 + 3 n\right)}\right]^{1/3\left(n - \omega\right)} - 1.
\end{equation} 
 For the case where DE is a cosmological constant $(\omega = - 1)$ and $\Omega_{0}^{(M)} = 0.3$ we get
 
\[
\bar{z} = \left(0.671, \; 0.511, \; 0.411, \; 0.326\right), \;\;\;\mbox{for}\;\;\;n = \left(0, \; 0.1, \; 0.2, \; 1/3\right).
\]
We note that the redshift at $q = 0$ predicted by the CDM models is a little more than twice the one obtained for $n = 1/3$. 
As a second example we consider phantom DE with $\omega = - 1.5$ and, once again, take $\Omega_{0}^{M} = 0.3$. We find 
\[
\bar{z} = \left(0.595, \;0.484, \;0.407, \; 0.337, \; 0.237, \; 0.207\right), \;\;\;\mbox{for}\;\;\;n = \left(0, \; 0.1, \; 0.2, \; 1/3, \; 2/3, \; 1\right). 
\]

Thus, for non-phantom ($\omega \geq  - 1$) and phantom matter $(\omega < - 1)$, we can conclude that the stiffer the primordial EoS, the later - i.e. closer to our era - the transition to accelerated expansion. 

\medskip

To finish the discussion, we would like to point out  that the asymptotic model  discussed here 
and the CDM model - given by (\ref{CDM solution}) - do not challenge each other. Instead, they are complementary.  Indeed, they represent two different limits: in the CDM model dust is assumed during the whole evolution, while here there is a sharp transition from a primordial $n \neq 0$ stage,  to cosmological dust.  These two asymptotic models should serve to restrict or restrain more realistic ones.

\subsection{Features of the model that are independent of $\alpha$ and $K$}

In any physical model is important to identify  the features that are independent of the particular choice of the parameters of the theory. Therefore, we now look for relationships between the observational parameters that are independent of the particular choice of $\alpha$ and $K$.  With this in mind we recast (\ref{x0 with n}) into the form

\[
X^{3\omega}\left[1 + \frac{1}{X}\right]^{3 n} = K^{3\omega}\left[1 + \frac{1}{K}\right]^{3 n} \left[\frac{\Omega_{0}^{(M)}}{\Omega_{0}^{(DE)}}\right]^{\alpha},
\]
where $X = K x^{\alpha}$. Since, $\alpha> 0$, $n \geq 0$ and $\omega < 0$, from this equation it follows that  $X > K$ for $\Omega_{0}^{(M)} < \Omega_{0}^{(DE)}$; 
$X < K$ for $\Omega_{0}^{(M)} > \Omega_{0}^{(DE)}$, and $X = K$ for $\Omega_{0}^{(M)} = \Omega_{0}^{(DE)}$. This is consistent with  the fact that  $x_{0} > 1$ ($x_{0} \leq 1$) $\Longleftrightarrow$ $\Omega_{0}^{M} < 1/2$ ($\Omega_{0}^{M} \geq 1/2$) discussed in (\ref{x0 for different Omegas}).

Now, from (\ref{q0 for the model with n}) we obtain $(n \neq 0)$  
\begin{equation}
\label{x0 for the model with n}
K x_{0}^{\alpha} = \frac{3 \left(\omega - n\right) \Omega_{0}^{(M)} + 2 q_{0} - 1 - 3\omega}{3\omega \Omega_{0}^{(DE)} - 2 q_{0} + 1}.
\end{equation}
Using this expression we find:

\paragraph{Models with $\Omega_{0}^{(M)} < 1/2$:} Setting $K x_{0} > K$, from (\ref{x0 for the model with n}) we get
\begin{equation}
\label{inequality for q, model with n}
\frac{1}{2}\left[1 + 3\omega\Omega_{0}^{(DE)}\right] < q_{0} < \frac{1}{2}\left[1 + 3\omega\Omega_{0}^{(DE)}\right] + \frac{3}{4}n \Omega_{0}^{(M)}.
\end{equation}
 As in the CDM models, the current accelerated expansion requires
\begin{equation}
\label{requirement for accelerated expansion, models with n}
\omega < - \frac{1}{3\Omega_{0}^{(DE)}}. 
\end{equation}
Equivalently, in terms of $\omega$ the above inequality can be written as
\begin{equation}
\label{equivalent inequality}
\left[\frac{2 q_{0} - 1}{3\Omega_{0}^{(DE)}}\right] - \frac{n \Omega_{0}^{(M)}}{2\Omega_{0}^{(DE)}}  <   \omega < \left[\frac{2 q_{0} - 1}{3\Omega_{0}^{(DE)}}\right]
\end{equation}

$\bullet$ For any given $\Omega_{0}^{(M)} < 1/2$, this expression yields a lower and/or an upper bound on $q_{0}$ for various values 
of $\omega$,  viz.,  
\begin{eqnarray}
\label{specific bounds on q}
q_{0} &>& - 1 + \frac{3\Omega_{0}^{(M)}}{2}, \;\;\;\mbox{for}\;\;\;\omega > - 1,\nonumber  \\
q_{0} &<& - 1 + \frac{3\left(n + 2\right)\Omega_{0}^{(M)}}{4}, \;\;\;\mbox{for}\;\;\; \omega < - 1,\nonumber \\
- 1 + \frac{3\Omega_{0}^{(M)}}{2} < q_{0} &<& - 1 + \frac{3\left(n + 2\right)\Omega_{0}^{(M)}}{4}, \;\;\;\mbox{for}\;\;\;\omega = - 1.
\end{eqnarray}
As an illustration, let us take  $\Omega_{0}^{(M)} = 0.3$. Then,  accelerated expansion requires $\omega < - 0.476$;  if  
$\omega  > - 1$ then the deceleration parameter has a  lower bound $q_{0}^{(min)} = - 0.550$, regardless of $n$;  if $\omega < - 1$  it has an upper bound that depends on $n$, namely,  $q_{0}^{(max)} = (- 0.325, \; - 0.447, \; -0.550)$ for $n = (1, \; 1/3, \; 0)$, respectively; if $\omega = - 1$ then $q_{0}$ is bounded from above and from bellow, viz.,  $- 0.550 < q_{0} < (- 0.325, \; - 0.447, \; -0.550)$ for $n = (1, \; 1/3, \; 0)$, respectively.

\paragraph{Models with $\Omega_{0}^{(M)} = 1/2$:} In these models the condition $x_{0} = 1$ yields

\begin{eqnarray}
\label{q for Omega = 1/2 with n}
q_{0} = \frac{1}{2}\left[1 + \frac{3}{4}\left(n + 2 \omega\right)\right].
\end{eqnarray}
Accelerated expansion at the current epoch requires
\begin{equation}
\omega <  - \frac{4 + 3 n}{6}.
\end{equation}

\paragraph{Models with $\Omega_{0}^{(M)} > 1/2$:} 
In these models the lower and upper bounds are generated by the condition $0 < x_{0} < 1$. Using (\ref{x0 for the model with n})  we find\footnote{$x_{0} > 0$ implies $\left[1 + 3\omega \Omega_{0}^{(DE)}\right]<  2 q_{0} < \left[ 1 + 3\omega\Omega_{0}^{(DE)}  + 3 n \Omega_{0}^{(M)}\right]$.}

\begin{equation}
\label{allowed range for q, Omega > 1/2}
 \frac{1}{2}\left[1 + 3 {\omega}\Omega_{0}^{(DE)}\right] +  \frac{3}{4} n \Omega_{0}^{(M)} < q_{0} < \frac{1}{2}\left[1 + 3 {\omega}\Omega_{0}^{(DE)}\right] + \frac{3}{2}n \Omega_{0}^{(M)}. 
\end{equation} 
This expression, together with $\Omega_{0}^{(M)} > 1/2$,  can be used to obtain more specific bounds  on $q_{0}$, similar to those in (\ref{specific bounds on q}).

Although observations indicate that the universe today is DE-dominated, it is of theoretical interest to note that  accelerated expansion may still occur in a matter-dominated phase if $|\omega|$ is large enough. Specifically, 

\begin{equation}
\label{q = 0 transition for matter dominated universe}
\omega <  -  \frac{\left[2 + 3 n \, \Omega_{0}^{(M)}\right]}{6\Omega_{0}^{(DE)}}.
\end{equation}
 As an illustration, for $\Omega_{0}^{(M)} = 0.6$ the transition from decelerated to accelerated expansion occurs in a recent past if $\omega < - \left(5/6 + 3 n/4\right) = - \left(0.833, \, 1.083, \, 1.583\right)$, for $n = \left(0,\, 1/3,\, 1\right)$, respectively. If $\omega > - \left(5/6 + 3 n/4\right)$, due to the continuous expansion of the universe, the transition occurs in the future, i.e., at some $- 1 < \bar{z} < 0.$ In the above expressions setting $n = 0$, we recover  the CDM model.

\section{Dark energy  with variable EoS and CDM ($w = 0$)}

In this section we consider the case where the DE component evolves according to the EoS (\ref{3.2}). Our aim is to probe the extent to which a dynamical EoS for DE  can affect and/or modify the straightforward description  of accelerated expansion as given by the $W = \omega = $ constant models. As a framework for  our discussion we consider a flat universe whose matter content is approximated by non-relativistic dust.  

 Thus, in this section we set $n = 0$, and keep  $\omega \neq \gamma$.  With this simplification (\ref{definition of x0}) reduces to 
\begin{equation}
\label{definition of x0 variable W}
x_{0}^{3 {\omega}}
\left(1 + \frac{1}{x_{0}^{\beta}}\right)^{3({\omega} - \gamma)/\beta} = 2^{3 \left(\omega - \gamma\right)/\beta}\left[\frac{\Omega_{0}^{(M)}}{\Omega_{0}^{(DE)}}\right].
\end{equation}
From (\ref{general expression for q in our problem}), with $n = 0$,  we get the equation for $\bar{x}$, viz.,

\begin{equation}
\label{equation for xbar with variable W}
{\bar{x}}^{3 \omega} + 2^{3\left(\omega - \gamma\right)/\beta}\left[1 + \frac{1}{{\bar{x}}^{\beta}}\right]^{- 3\left(\omega - \gamma\right)/\beta}
\left[\frac{\left(1 + 3 \omega\right) {\bar{x}}^{\beta} + \left(1 + 3 \gamma\right)}{{\bar{x}}^{\beta} + 1}\right] = 0.
\end{equation}
For $\gamma = \omega$ we recover the CDM model (\ref{CDM solution}).

\medskip

As we mentioned in Section (3.2), the EoS (\ref{3.2}) - although in a slightly different  notation - has been studied by Hannestad and M\"{o}rtsell in  Ref. \cite{EoS}. From a joint analysis of data from the cosmic microwave
background, large scale structure and type-Ia supernovae, these authors evaluated the  constraints on (\ref{3.2}) and concluded that the best fit model corresponds, in our notation,  to ${\omega} = - 1.8$, $\gamma = - 0.4$, $\beta = 3.41$ and $ \Omega_{0}^{(M)} = 0.38$.
For these values from (\ref{definition of x0 variable W})-(\ref{equation for xbar with variable W}) we get $x_{0} = 1.148$ and $\bar{x} = 0.789$. Accordingly, the universe becomes DE-dominated at $z_{*} = 0.148$ and starts accelerating expansion at $\bar{z} = 0.452$. From (\ref{q today}) we find  $q_{0} = - 0.673$ and $W_{0} = - 1.262$.

\medskip

However, the observational facts are not yet conclusive as to settle the  question of the dynamics in the dark energy EoS. Indeed there are  other choices of parameters that lead  to results consistent with 
observational constrains. Therefore, here we concentrate our attention on studying the different possibilities and models for DE offered and/or suggested  by (\ref{3.2}). 

\medskip

A short inspection of (\ref{definition of x0 variable W})-(\ref{equation for xbar with variable W}) suggests we start with the examination of three different cases: (i) $\beta = 3 \left(\gamma - \omega\right)$; (ii) $0 <\beta \ll 3 \left(\gamma - \omega\right)$ and (iii) $\beta \gg 3 \left(\gamma - \omega\right)$. The first case corresponds to the superposition of fluids mentioned in (\ref{density and pressure, superposition of fluids}). The second one yields the CDM solution (\ref{CDM solution}) with $\omega$ replaced by $\left(\omega + \gamma\right)/2$. The third case characterizes the limiting case 
where $W$ rapidly evolves from $\gamma$ to  $\omega$.

\subsection{Quick transition from $W = \gamma$ to $W = \omega$}

We now proceed to study the properties of the limiting model. For  $\beta \gg 1$ the solution to (\ref{definition of x0 variable W}) is given by

\begin{equation}
\label{solution for x0 and different Omegas, 2}
 x_{0} =  \left\{\begin{array}{cc}
 \left[\frac{\Omega_{0}^{(M)}}{\Omega_{0}^{(DE)}}\right]^{1/3\omega}& \mbox{for $\Omega_{0}^{(M)} < 1/2$}, \\
\left[\frac{\Omega_{0}^{(M)}}{\Omega_{0}^{(DE)}}\right]^{1/3 \gamma} & \mbox{for $\Omega_{0}^{(M)} > 1/2$}, 
 \end{array}
\right.
\end{equation}
where we have assumed $\gamma \neq 0$ and $\omega \neq 0$. The solutions with $\gamma = 0$ and $\omega = 0$ will be discussed bellow.

Regarding (\ref{equation for xbar with variable W}), for $\beta \gg 1$ we find several solutions depending on whether $\omega$ and $\gamma$ are greater or less than $-2 / 3$. These are\footnote{If $\bar{x} > 1$, then ${\bar{x}}^{\beta} \gg 1$ for $\beta \gg1$. Consequently, in this limit $\bar{x} = \left[- \left(1 + 3 \omega\right)\right]^{1/3\omega}$.  For consistency, the requirement $\bar{x} > 1$ demands $\omega > - 2/3$.  
If $\bar{x} < 1$, then ${\bar{x}}^{\beta} \ll 1$ for $\beta \gg1$. Thus, in this limit $\bar{x} = \left[- \left(1 + 3 \gamma\right)\right]^{1/3\gamma}$.  The requirement $\bar{x} < 1$ demands $\gamma < - 2/3$.}
:

\paragraph{(i)} For $\omega > - 2/3$, $\gamma \geq - 2/3$ the solution is $\bar{x} = \left[- \left(1 + 3 \omega\right)\right]^{1/3\omega}$. The requirement $\omega > - 2/3$ asserts that $\bar{x} > 1$.

\paragraph{(ii)} For $\omega > - 2/3$, $\gamma < - 2/3$ there are three distinct roots: ${\bar{x}}_{1} = \left[- \left(1 + 3 \gamma\right)\right]^{1/ 3\gamma}$; one of the following 
 ${\bar{x}}_{2} = \left(1^{+}, \; 1, \; 1^{-}\right)$, depending on whether $q_{*} < 0$, $q_{*} = 0$ or $q_{*} > 0$, respectively; and  
${\bar{x}}_{3} = \left[- \left(1 + 3 \omega\right)\right]^{1/3\omega}$. The conditions on $\omega$ and $\gamma$ establish that  ${\bar{x}}_{1} < 1$,  ${\bar{x}}_{3} > 1$. Also, here  
\[
q_{*} = q (1) = \frac{1}{2} + \frac{3}{8}\left(\omega + \gamma\right).
\]
Thus, for a given $\omega$ the statement  $q_{*} \geq  0$  denotes $\gamma \geq  - \left(\omega + 4/3\right)$, and vice versa. 

\paragraph{(iii)} For $\omega \leq - 2/3$, $\gamma \geq - 2/3$ we find either $\bar{x} =  1^{+}$ or $\bar{x} =  1^{-}$ depending on whether $q_{*} >  0$ or $q_{*} <  0$, respectively.

\paragraph{(iv)} For $\omega \leq - 2/3$, $\gamma < - 2/3$ the solution is  $\bar{x} = \left[- \left(1 + 3 \gamma\right)\right]^{1/ 3\gamma} $. The requirement $\gamma < - 2/3$ asserts that $\bar{x} < 1$.

\paragraph{(v)} Finally, when $q_{*} = 0$ we find $\bar{x} = 1$ regardless of $\omega, \gamma$ and $\beta$, as expected.

\bigskip

In Table $3$ we illustrate the solutions to (\ref{equation for xbar with variable W}). They should be compared with the CDM with $\omega = \gamma$, which are given by the first column in Table $2$. They drastically differ for $\omega < - 2/3$. 

\bigskip

\begin{center}
\begin{tabular}{|c|c|c|c|c|c|} \hline
\multicolumn{6}{|c|}{\bf Table 3. $\bar{x}$ for various $\omega$, $\gamma$ and $\beta \gg 1$}\\ \hline
 $ \omega$ & $ \gamma = - 0.4$& $\gamma = - 2/3$& $ \gamma = - 5/6$ &$\gamma = - 1$&$ \gamma = - 1.5$    \\ \hline
$ - 0.4$ & $3.824$& $3.824$& $ 0.850, \; 1^{-}, \; 3.824$ &$0.794, \;1^{+}, \; 3.824$ &$0.757, \; 1^{+}, \; 3.824$   \\ \hline
$ - 0.5$ & $1.587$& $1.587$& $0.850, \; 1, \; 1.587$ &$0.794, \;1^{+}, \; 1.587$ &$0.757, \; 1^{+}, \; 1.587$   \\ \hline
$ - 2/3$ & $1^{+}$& $1$& $0.850$ &$0.794$ &$0.757$   \\ \hline
$ - 1$ & $1^{-}$& $1^{-}$& $0.850$ &$0.794$ &$0.757$  \\ \hline
$ - 1.5$ & $1^{-}$& $1^{-}$& $0.850$ &$0.794$ &$0.757$   \\ \hline
\end{tabular}
\end{center}

\bigskip

 In terms of the cosmological redshift the solutions to (\ref{equation for xbar with variable W}) are
\begin{eqnarray}
{\bar{z}}_{\omega} &=& \frac{\left[\frac{\Omega_{0}^{(M)}}{\Omega_{0}^{(DE)}}\right]^{1/3 \omega}}{\left[- \left(1 + 3 \omega\right) \right]^{1/3 \omega}} - 1, \nonumber \\
{\bar{z}}_{*} &=& \left[\frac{\Omega_{0}^{(M)}}{\Omega_{0}^{(DE)}}\right]^{1/3 \omega} - 1, \nonumber \\
{\bar{z}}_{\gamma} &=& \frac{\left[\frac{\Omega_{0}^{(M)}}{\Omega_{0}^{(DE)}}\right]^{1/3 \omega}}{\left[- \left(1 + 3 \gamma\right) \right]^{1/3 \gamma}} - 1.
\end{eqnarray}
Solution (i) generates models that coincide with a  subset of the $\gamma = \omega$ models given by (\ref{CDM solution}) where $q$ crosses from plus to minus at $z = {\bar{z}}_{\omega}$. 
Solutions (ii)-(iv) show that an evolving  DE with $\omega \neq \gamma$ may affect not only the epoch at which $q$ vanishes, but also the 
 number of  times it changes its sign. 

Models with $\omega$ and $\gamma$ in the range given by (ii) have $d W/d x > 0$ and $q$ goes through zero three times: the first change is from plus to minus at $z = {\bar{z}}_{\gamma}$. The next change is from minus to plus at $z = {\bar{z}}_{*}$. Finally, there is another transition from plus to minus at $z = {\bar{z}}_{\omega}$. As an illustration, let us take $\omega = - 0.5$, $\gamma = - 2/3$ and $\Omega_{0}^{(M)} = 0.3$. From Table $1$ we get $x_{0} = 1.759$. Using the numbers given in Table $3$ we get ${\bar{z}}_{\gamma} = 1.069$, ${\bar{z}}_{*} = 0.759$, ${\bar{z}}_{\omega} = 0.10$.  These solutions are similar to the one illustrated in Fig. \ref{fig:q0}.

 Models constructed from solutions (iii) have $d W/d x < 0$ and $q$ goes through zero only once, from plus to minus. The remarkable  feature here is that the transition is at $\bar{x} \approx 1$ for {\it{all}} $\omega < 2/3$ and $\gamma \geq 2/3$. However the observable redshift of transition $z = {\bar{z}}_{*}$ does depend on $\omega$. 

 Models generated by solution  (iv) are interesting because the transition from positive to negative $q$ depends only on $\gamma$, so it occurs at the same $\bar{x}$ for all $\omega < - 2/3$. But,  the corresponding redshift  $z = {\bar{z}}_{\gamma}$ does explicitly depend on $\omega$.

\subsection{Solution for $\gamma = 0$: DE evolving from non-relativistic dust}

\begin{figure}[tbp] 
  \centering
  \includegraphics[width=3.00in,height=3.00in,keepaspectratio]{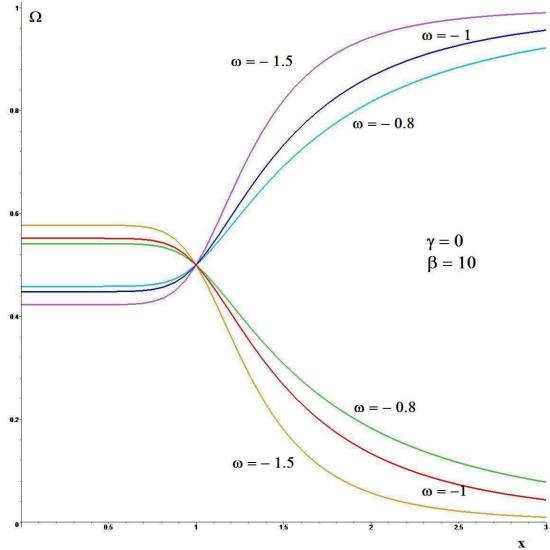}
  \caption{The decreasing functions represent the behavior of  $\Omega^{(M)}$ for models with $\gamma = 0$ and various choices of $\omega$.  They  decrease from $\Omega^{(M)} = \Omega^{(M)}_{max} = \left[1 + 2^{3\omega/\beta}\right]^{- 1}$ at early times to  $\Omega^{(M)} = 0 $ as $x \rightarrow \infty$ at late times. Clearly,  $1/2 < \Omega^{(M)}_{max} <  1$ for any $\omega < 0$ and $\beta > 0$. Keeping  $\omega$ fixed we find that $\Omega^{(M)}_{max} \rightarrow (1/2)^{+}$ for $\beta \gg 1$ and $\Omega^{(M)}_{max} \rightarrow 1^{-}$ for $\beta \rightarrow 0$. In the case under consideration, for $\beta = 0$ we recover the usual CDM model with $W = \omega/2$. The increasing functions give  $\Omega^{(DE)}$.
}
  \label{fig:gammazerOmega}
\end{figure}

In (\ref{solution for x0 and different Omegas, 2}) we have assumed $\gamma \neq 0$. However models with $\gamma = 0$ are interesting because - following our discussion at the end of section $(3.2)$ - they behave like pressureless dust at early times when $x$ is small and as DE with $p^{(DE)} = \omega \rho^{(DE)}$ at late times. It turns out that (\ref{definition of x0 variable W}) for $\gamma = 0$ admits a nice solution, viz.,  

\begin{equation}
\label{x0 for generalized Chaplygin gas}
x_{0} = \left[ 2\left(\frac{\Omega_{0}^{(DE)}}{\Omega_{0}^{(M)}}\right)^{- \beta/3 \omega} - 1\right]^{1/\beta}.
\end{equation}
Since the quantity inside the brackets must be positive we should require (See Figure \ref{fig:gammazerOmega})

\begin{equation}
\label{condition for solution with gamma zero}
\Omega_{0}^{(M)} < \frac{1}{1 + 2^{3 \omega/\beta}}.
\end{equation}

At the current epoch the EoS  and deceleration parameter are given by

\[
W_{0} = \omega \left[1 - \frac{1}{2}\left(\frac{\Omega_{0}^{(M)}}{\Omega_{0}^{(DE)}}\right)^{- \beta/3 \omega}\right]
\]
\[
q_{0} = \frac{1}{2}\left[1 + 3 \Omega_{0}^{(DE)} W_{0}\right]
\]
Thus, $W_{0} \rightarrow \omega$ and $q_{0} \rightarrow \left(1 + 3 \omega\right)/2$ as $\Omega_{0}^{(DE)} \rightarrow 1$.

\begin{figure}[tbp] 
  \centering
  \includegraphics[width=3.00in,height=3.00in,keepaspectratio]{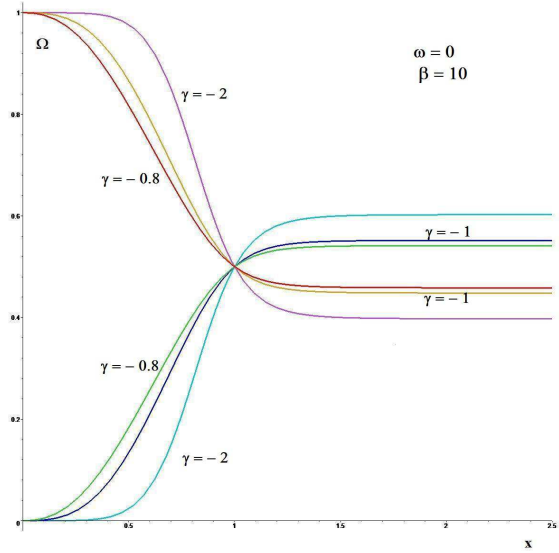}
  \caption{The decreasing functions represent the behavior of  $\Omega^{(M)}$ for models with $\omega = 0$ and various choices of $\gamma$. The increasing functions give  $\Omega^{(DE)}$. As the universe expands $\Omega^{(M)} \rightarrow 0$ when $\omega \neq 0$. However, $\Omega^{(M)} \rightarrow \Omega^{(M)}_{min} = \left[1 + 2^{- 3\gamma/\beta}\right]^{- 1}$ when $\omega = 0$. Clearly,  $0 < \Omega^{(M)}_{min} <  1/2$ for any $\gamma < 0$ and $\beta > 0$.  Besides, $\Omega^{(M)}_{min} \rightarrow (1/2)^{-}$ for $\beta \gg 1$ and $\Omega^{(M)}_{min} \rightarrow 0$ for $\beta \rightarrow 0$. In the case under consideration, for $\beta = 0$ we recover the usual CDM model with DE characterized by a constant EoS, viz.,  $W = \gamma/2$.}
\label{fig:omegazerOmega}
\end{figure}

\begin{figure}[tbp] 
  \centering
  \includegraphics[width=3.00in,height=3.00in,keepaspectratio]{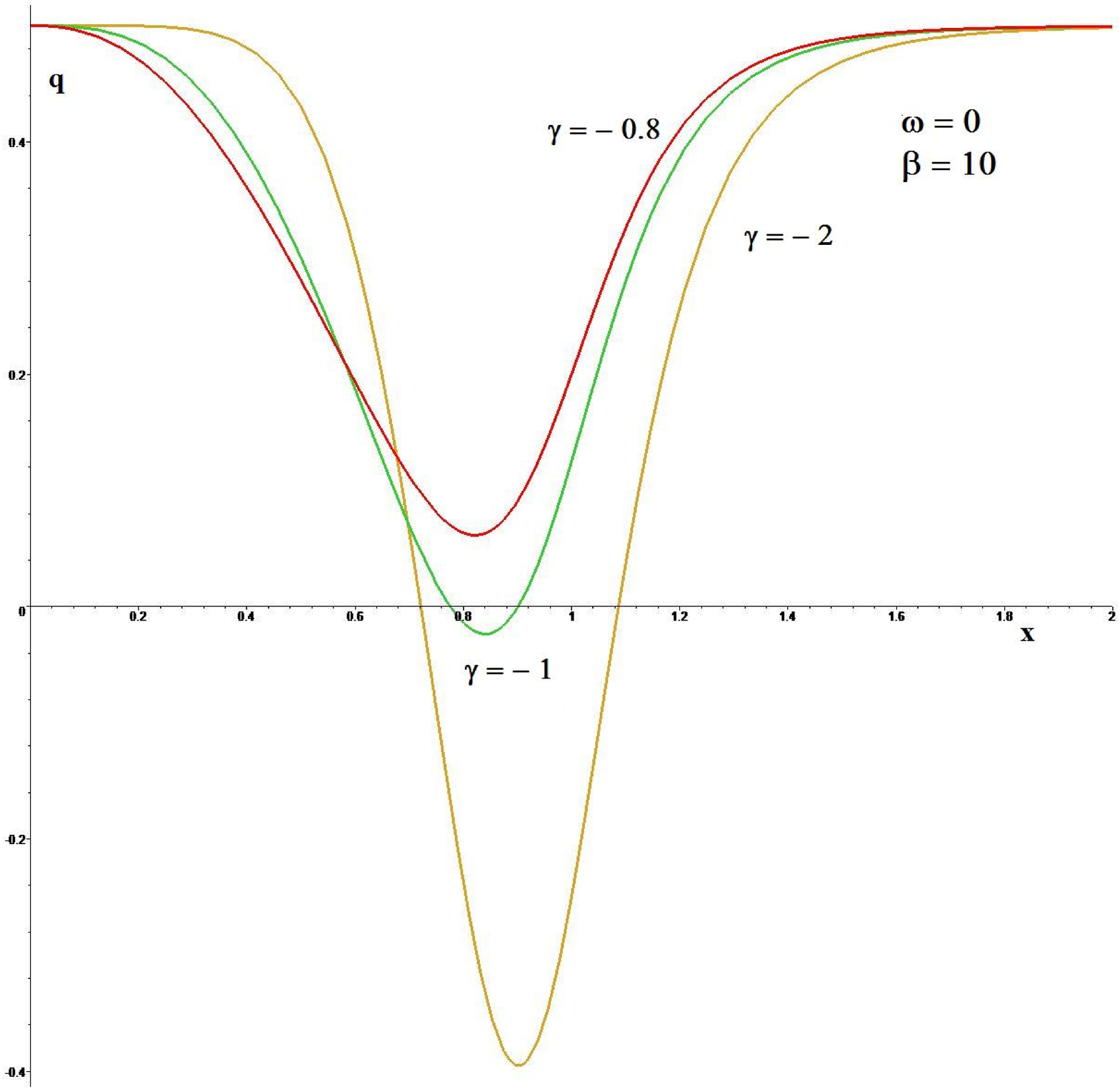}
  \caption{The figure illustrates that accelerated expansion may occur in models with $\omega = 0$ for different choices of $\gamma$. The deceleration parameter  has a minimum, say $q_{min}$,  around the transition epoch $x = 1$. For a fixed $\beta$ the minimum is negative or positive depending on $\gamma$. In the Figure $\beta = 10$. For this value $q_{min}$ is negative if $\gamma < - 0.943$.  In general,  the larger the magnitude of $\gamma$,  the greater the magnitude of $q_{min}$ .  Also the accelerated expansion starts earlier and finishes later with the increase of $|\gamma|$.}
  \label{fig:omegazero1}
\end{figure}

\paragraph{Chaplygin gas:}  If in addition to $\gamma = 0$ we set $\omega = - 1$ then the  energy density and pressure of the DE component satisfy the EoS (\ref{EoS for Chaplygin gas}). When $\beta$ changes in the range $3 \leq \beta \leq 6$ we recover the so-called generalized Chaplygin gas. For $\beta = 6$, which corresponds to the original Chapligin gas,   the solution to (\ref{equation for xbar with variable W}) is $\bar{x} = 1.069$. To obtain specific numbers for the redshifts we need to feed the equations 
with $\Omega_{0}^{(M)}$. The model requires $\Omega_{0}^{(M)} < 2 - \sqrt{2} \approx 0.586$. Thus,  as an illustration we 
take $\Omega_{0}^{(M)}  =0.3$. With this choice we find $x_{0} = 1.465$, $W_{0} = - 0.908$ and $q_{0} = - 0.453$. The universe becomes dominated by the gas at $z_{*} = 0.465$ and the phase of accelerated expansion  starts at $\bar{z} = 0.370$.

\subsection{Solution for $\omega = 0$: dark energy transforming into non-relativistic cosmic dust}

Thus far we have assumed $\omega \neq 0$. Models with $\omega = 0$ are diametrically opposite to the ones with $\gamma = 0$. They behave as DE with $p^{(DE)} = \gamma  \rho^{(DE)}$ at early times and like pressureless dust at late times. Immediately after (\ref{asymptotic behavior of omegade}), we mentioned that the replacement  $\beta \rightarrow - \beta$ changes the asymptotic role of $\omega$ and $\gamma$. The solution to (\ref{definition of x0 variable W}) is now given by

\begin{equation}
x_{0} = \left[2\left(\frac{\Omega_{0}^{(M)}}{\Omega_{0}^{(DE)}}\right)^{- \beta/3\gamma} - 1\right]^{- 1/\beta},
\end{equation}
which can readily be obtained from (\ref{x0 for generalized Chaplygin gas}) by replacing $\beta \rightarrow - \beta$ and $\omega \rightarrow \gamma$. However, the analogy is not complete; namely instead of (\ref{condition for solution with gamma zero}) - with the corresponding changes -  in the case under consideration the condition on $\Omega_{0}^{(M)}$ is reversed, viz.,

\[
\Omega_{0}^{(M)} > \frac{1}{1 + 2^{- 3 \gamma/\beta}}.
\]
Since $\beta > 0$, $x_{0} \rightarrow \infty$ as $\Omega_{0}^{(M)}\rightarrow \left[1 + 2^{- 3\gamma/\beta} \right]^{- 1}$ and $q \rightarrow 1/2$ as corresponds to dust models. In Figure \ref{fig:omegazerOmega} we illustrate this for various choices of $\gamma$.

\begin{figure}[tbp] 
  \centering
  \includegraphics[width=3.00in,height=3.00in,keepaspectratio]{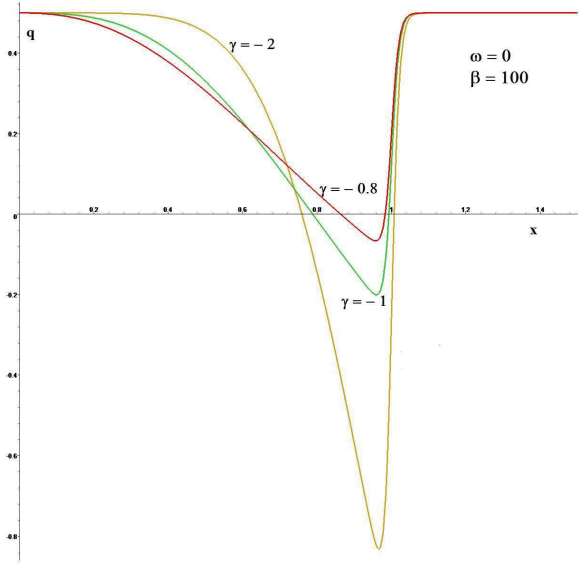}
  \caption{The figure illustrates the role of $\beta$, which measures the rapidity of transition from $\omega$ to $\gamma$. The larger $\beta$ the shorter the transition. Comparison with 
Figure \ref{fig:omegazero1} illustrates  that  increasing $\beta$  decreases the duration of the accelerated expansion and raises the magnitude of $q_{min}$. In this respect, the effects of $\gamma $ and $\beta$ are very much alike. This is  because, when $\omega = 0$,  both affect the slope of $W$ in a similar way, viz., $d W/ d x \sim - \gamma \beta/4$ near $x = 1$.}
  \label{fig:omegazero2}
\end{figure}

The question of whether or not this type of DE leads to an accelerated expansion depends on the choice of $\gamma$ and $\beta$. This is illustrated in Figures \ref{fig:omegazero1} and \ref{fig:omegazero2}.

\paragraph{Reversed Chaplygin gas:} For the particular choice  $\gamma = - 1$ we have  

\[
p^{(DE)}\left[\rho^{(DE)}\right]^{\lambda} = - C_{2}^{\lambda + 1}, \;\;\;\;\mbox{with}\;\;\;\;\lambda = - \frac{3  + \beta }{3},
\]
and 
\[
\rho^{(DE)} = \frac{C_{2}}{\left(x^{\beta} + 1\right)^{3/\beta}}, \;\;\;\;p = - \frac{C_{2}}{\left(x^{\beta}+ 1\right)^{\left(3 + \beta\right)/\beta}}.
\]
For  $\beta  > 0$, the fluid behaves like a cosmological constant for $x  \ll 1$ and dust at late times. Matter dominates for small $x$ so that $q = 1/2$ at early times and at late times. The expansion becomes accelerated in between for  $\beta > 9$. See Figures  \ref{fig:omegazerOmega}, \ref{fig:omegazero1}, \ref{fig:omegazero2}.

\subsection{DE acting as a cosmological constant at late times: $\omega = - 1$ }

To obtain an explicit expression relating $\rho^{(DE)}$ and $p^{(DE)}$ we use (\ref{3.2}) to isolate $x$ and then substitute the result into (\ref{3.5}). 
To ensure that the DE behaves as a cosmological constant at late times we set $\omega = -1$. Solving  for $p^{(DE)}$ we obtain

\begin{equation}
\label{DE acting as CC}
p^{(DE)} = \rho^{(DE)} \left[\gamma - \left(\gamma + 1\right)\left(\frac{\rho^{(DE)}}{C_{2}}\right)^{- \beta/3\left(\gamma + 1\right)}\right], \;\;\;\;\gamma \neq - 1. 
\end{equation}
This can be interpreted as the superposition of two non-interacting fluids, with energy densities $\rho_{1}$ and $\rho_{2}$,  evolving according to the EoS
\begin{equation}
p_{1} = \gamma \rho_{1}, \;\;\;\;p_{2} = - A \rho^{[1 - \beta/3\left(\gamma + 1\right)]},
\end{equation}
where $A = \left(\gamma + 1\right) C_{2}^{\beta/3\left(\gamma + 1\right)}$. Thus, from the continuity equation we get
\begin{equation}
\label{densities for DE acting a CC}
\rho_{1} = \frac{B_{1}}{x^{3\left(1 + \gamma\right)}}, \;\;\;\;\rho_{2} = \left[A + \frac{B_{2}}{x^{\beta/\left(1 + \gamma\right)}}\right]^{3\left(1 + \gamma\right)/\beta},
\end{equation}
where $B_{1}$ and $B_{2}$ are constants of integration. The above expression requires $\gamma > - 1$, i.e., $A > 0$, otherwise $\rho_{2}$ diverges at some finite $x \neq 0$. For $\gamma > - 1$,  $\rho_{1}$ acts as  quintessence and $\rho_{2}$ at early times - when $x$ is small -  behaves as pressureless dust: $\rho_{2}\approx 1/x^3$. At late times $\rho_{2}$ behaves as a cosmological term since $p_{2} = - \rho_{2}$ in that limit. For $\gamma = 0$ and $\beta = 6$ we have a combination of dust and Chaplygin gas.

It is well known that a scalar field  $\phi$ minimally coupled to gravity with a potential $V(\phi)$ can serve as a model for DE (see e.g. \cite{Review}).  In a DE-dominated universe a first order formalism gives $\left(d \phi/d x\right)^2 = 3\epsilon \left(1 + W\right)/x^2$ and $V = \left(1 - W\right)\rho/2$, where $\epsilon = 1$ or $\epsilon = - 1$ depending  on whether $\phi$ is an ordinary scalar field ($W > - 1$) or a phantom field ($W < - 1$).

For $\omega = - 1$ we get 
\[
x^{\beta} = \frac{1}{\sinh^2 \eta \phi}, \;\;\;\eta = \frac{\beta}{2 \sqrt{3}\sqrt{1 + \gamma}}.
\]

\begin{equation}
\label{potential for omega = - 1}
V(\phi) = \frac{C_{2}}{2}\left[\left(\cosh \eta\phi\right)^{6\left(\gamma + 1\right)/\beta} + \left(1 - \gamma \sinh^2 \eta\phi\right)\left(\cosh \eta \phi\right)^{2\left[3\left(\gamma + 1\right) - \beta\right]/\beta}\right].
\end{equation}
For $\gamma = 0$ and $\beta = 6$ this expression gives the potential for the Chaplygin gas discussed in \cite{Review}.

\subsection{DE acting as non-relativistic cosmological dust  at late times: $\gamma = - 1$}

Following the same steps leading to (\ref{DE acting as CC}), but this time setting $\gamma = - 1$ we obtain

\begin{equation}
\label{DE acting as CC t early times}
p^{(DE)} = \rho^{(DE)} \left[\omega - \left(\omega + 1\right)\left(\frac{\rho^{(DE)}}{C_{2}}\right)^{\beta/3\left(\omega + 1\right)}\right], \;\;\;\;\omega \neq - 1. 
\end{equation}
Once again the DE component can be interpreted as the superposition of two non-interacting fluids, viz.,
\[
p_{1} = \omega \rho_{1}, \;\;\;\;\;p_{2} = - \tilde{A}\rho_{2}^{\left[1 + \beta/3\left(\omega + 1\right)\right]}, 
\]
with $\tilde{A} = \left(\omega + 1\right) C_{2}^{- \beta/3\left(\omega + 1\right)}$ and 
\[
\rho_{1} = \frac{{\tilde{B}}_{1}}{x^{3\left(1 + \omega\right)}}, \;\;\;\;\;\rho_{2} = \frac{1}{x^3}\left[{\tilde{B}}_{2} +
 \frac{\tilde{A}}{x^{\beta/\left(1 + \omega\right)}}\right]^{- 3\left(1 + \omega\right)/\beta}, 
\]
where ${\tilde{B}}_{1}$ and ${\tilde{B}}_{2}$ are constants of integration, and $\tilde{A} >0$ to avoid singularities at some finite $x$. In this case the evolution of $\rho_{2}$  is diametrically opposed to that discussed in (\ref{DE acting as CC})-(\ref{densities for DE acting a CC}). Specifically, now $p_{2} = - \rho_{2} = {\tilde{A}}^{- 3\left(\omega + 1\right)/\beta}$ at early times and $\rho_{2} \approx 1/x^3$ at late times.  A particular example of this model is the case where $\omega = 0$, which corresponds to the reversed Chaplygin gas  discussed in Section $5.3$.  

Following the same procedure used in the preceding subsection, we can obtain a potential by treating DE with $\gamma = - 1$ as an ordinary scalar field. We find

\[
x^{\beta} = \sinh^2 {\bar{\eta}} \phi, \;\;\;\bar{\eta} = \frac{\beta}{2 \sqrt{3}\sqrt{1 + \omega}}.
\]
and 
\begin{equation}
\label{potential gamma = - 1}
V(\phi) = \frac{C_{2}}{2}\left[\left(\cosh \bar{\eta}\phi\right)^{- 6\left(1 + \omega\right)/\beta} + \left(1 - \omega \sinh^2\bar{\eta}\phi\right)\left(\cosh \bar{\eta}\phi\right)^{- 2\left[3\left(1 + \omega\right) + \beta\right]/\beta}\right].
\end{equation}

\medskip

In order to avoid a possible misunderstanding, we should mention that the models with $\gamma = 0$,  discussed in Section $5.2$,  {\it {cannot}} be expressed as  a superposition of fluids, except in the case where $\omega = - 1$. The same is true for the models with $\omega = 0$ of Section $5.3$, which  can be represented as a superposition of fluids only when $\gamma = - 1$.   

\subsection{Parameter space}

In the CDM models there is a simple relationship between the cosmological parameters at the current epoch. This is given by the third equation in (\ref{CDM solution}).  In the case under consideration,  we can combine (\ref{q today}), with $n = 0$,  and (\ref{definition of x0 variable W}) to eliminate $x_{0}$ and obtain a single equation, viz.,
  \begin{equation}
\label{single equation for parameters with beta}
\left[\frac{3 \left(\gamma - \omega\right)\Omega_{0}^{(DE)}}{1 - 2 q_{0} +  3 \gamma \Omega_{0}^{(DE)}}\right]^{- \gamma}
\left[\frac{2 q_{0} - 1 -  3 \omega \Omega_{0}^{(DE)}}{3 \left(\gamma - \omega\right)\Omega_{0}^{(DE)}}\right]^{ - \omega} = 2^{\left(\omega - \gamma\right)}\left[\frac{\Omega_{0}^{(M)}}{\Omega_{0}^{(DE)}}\right]^{\beta/3}.
\end{equation}
 This replaces (\ref{CDM solution}) when $\omega \neq \gamma$.

Apart from this, some general constraints on the current values of the parameters can be obtained from  (\ref{q today}).  Solving for $x^{\beta}_{0}$ we get ($\omega \neq \gamma$)

\begin{equation}
x^{\beta} = - \frac{3 \gamma \, \Omega_{0}^{(DE)} - 2 q_{0} + 1}{3 \omega\, \Omega_{0}^{(DE)} - 2 q_{0} + 1}.
\end{equation}
For $\Omega_{0}^{(M)} < 1/2$ the condition $x_{0} > 1$ gives:

$\bullet$ For $\left(\omega - \gamma\right) > 0$:
\begin{equation}
\label{inequality for omega larger than gamma}
\frac{1}{2} + \frac{3}{4}\left(\gamma + \omega\right) \Omega_{0}^{(DE)} < q_{0} < \frac{1}{2} + \frac{3}{2} \omega\, \Omega_{0}^{(DE)}.
\end{equation}
Accelerated expansion requires 
\begin{equation}
\left(\omega + \gamma\right) < - \frac{2}{3 \Omega_{0}^{(DE)}}.
\end{equation}
Since $\omega > \gamma$, it follows that $\gamma < - 1/3\Omega_{0}^{(DE)}$.

$\bullet$ For $\left(\omega - \gamma\right) < 0$:

\begin{equation}
\label{inequality for omega less than gamma}
\frac{1}{2} + \frac{3}{2} \omega\, \Omega_{0}^{(DE)} < q_{0} < \frac{1}{2} + \frac{3}{4}\left(\gamma + \omega\right) \Omega_{0}^{(DE)}.
\end{equation}
The r.h.s. part of this inequality, together with $\Omega_{0}^{(DE)} > 1/2$ and $\omega < \gamma$, leads to $\gamma > - 2 \left(1 - 2 q_{0}\right)/3$.

$\bullet$ It should be noted that the inequalities (\ref{inequality for omega larger than gamma}), (\ref{inequality for omega less than gamma}) guarantee the positivity of the quantities inside the brackets in (\ref{single equation for parameters with beta}) for $\omega > \gamma$ and $\omega < \gamma$, respectively.

\section{Summary and conclusions}

Models in science  perform  two fundamentally different - but complementary - tasks. On the one hand, a model can indicate the type of phenomena that could actually occur in the context of the theory. On the other hand, it may shed light upon some problems of the theory.

 The CDM models with $W = \omega = $ constant are important first approximations in the description of DE. They satisfactorily explain the observational data for small values of $z$, which seems to be reasonable since the equations of state of the
constituents of the universe have probably not changed much in recent times. The natural question to ask is whether these models will still be useful when we have access to observations of  higher values of $z$. One would expect that the details of the expansion of the universe, which are determined by the evolution of its components, will become increasingly important as we look into the past.

In this paper we have constructed a simple cosmological model which is physically reasonable, mathematically tractable, and extends the study of CDM models to the case where both the EoS for matter and DE vary with time. It is given by equations (\ref{H in terms of z})-(\ref{q general in terms of z}). The deviation from CDM models is measured by the functions $f(z)$ and $g(z)$ given by (\ref{definition of f and g in terms of z}). For CDM models $f(z) = g(z) = 1$. In our model  $f(z) \approx 1$, $g(z) \approx 1$  for small $z$ and  $f(z) \approx \left(1 + z\right)^{3 n}$, $g(z) \approx \left(1 + z\right)^{- 3\left(\omega - \gamma\right)}$ for large values of $z$. Consequently, the model constructed here approximates the CDM ones for small $z$ but significantly differ from them for large $z$.

We have done a thorough review of all possible cases in the whole range of parameters of the EoS (\ref{3.2}), (\ref{EoS with n})  and discussed the corresponding physical models. Some of them are quite exotic as, e.g., those with $\omega > - 2/3$ and $\gamma < - 2/3$ (see Table $3$) that go from decelerated to accelerated expansion, and vice versa, several times. An example is illustrated  in Fig \ref{fig:q0}. 

In section $5$ we investigated the effects of a primordial EoS on the present accelerated expansion. In the CDM model there is a simple relationship between the current values of the cosmological parameters $q, \omega, \Omega^{(DE)}$, which is given by (\ref{CDM solution}). When  $w \neq 0$ that simple equation is replaced by three inequalities, viz.,  (\ref{inequality for q, model with n}), (\ref{q for Omega = 1/2 with n}), (\ref{allowed range for q, Omega > 1/2}) for 
$\Omega_{0}^{(M)} < 1/2$,  $\Omega_{0}^{(M)} =  1/2$ and $\Omega_{0}^{(M)} >  1/2$, respectively. They explicitly depend on $n$ in such a way that setting $n = 0$ we recover the CDM expression (\ref{CDM solution}). 

We also studied the asymptotic model $\alpha \gg n$ which corresponds to the case where $w$ abruptly changes from $w = n$ to $w = 0$. For $\omega \geq - 2/3$ and $\Omega_{0}^{(M)} < 1/2$ this and the CDM model are indistinguishable in the sense that they give the same values for $x_{0}$, $\bar{x}$, $z_{*}$ and $\bar{z}$, regardless of $n$. The situation is quite different when $\omega < - 2/3$.  

We note that the solutions for $\bar{x}$ sharply depend on whether DE is phantom ($\omega < - 1$) or non-phantom ($\omega \geq - 1$).  Specifically, when the universe evolves from a radiation dominated early stage $(n = 1/3)$ from the solutions (ii)-(v) of section $5.2$ we find  that $\bar{x} = 1$ for  $- 1 \leq \omega < - 2/3$ and $\bar{x} = \left[- \left(1 + 3 \omega\right)/2\right]^{1/\left(3\omega - 1\right)}$ for  $\omega < - 1$. A similar situation is seen in (\ref{specific bounds on q}) where the limits on $q_{0}$ are clearly dependent on the nature of DE. 

To illustrate the early and late behavior of the model, we consider the density parameter 
\[
\Omega^{(M)} =  \frac{\left(K + 1\right)^{- 3 n/\alpha}\left(K x^{\alpha} + 1\right)^{3 n/\alpha}}{\left(K + 1\right)^{- 3 n/\alpha}\left(K x^{\alpha} + 1\right)^{3 n/\alpha} + x^{3\left(n - \gamma\right)} 2^{3\left(\omega - \gamma\right)/\beta}\left(x^{\beta} + 1\right)^{- 3\left(\omega - \gamma\right)/\beta}}.
\] 
Near the origin, for small values of $x$, there are three different cases: 
(i) $\Omega^{(M)} = 1$, (ii) $\Omega^{(M)}  < 1$ and (iii) $\Omega^{(M)} = 0$ corresponding to $\left(n - \gamma\right) > 0$, $\left(n - \gamma\right) = 0$ and $\left(n - \gamma\right) < 0$, respectively. The models discussed in section $5$ have $n \geq 0$ and $\omega = \gamma < 0$ so they illustrate case (i). The models studied in section $6.2$, which include the generalized Chaplygin gas, have $n = \gamma = 0$ so they  illustrate case (ii).  We have restricted our discussion to $n \geq 0$, $\gamma \leq 0$ so case (iii) has been excluded from consideration. 
In the limit $x \rightarrow \infty$ there are two possibilities: $\Omega^{(M)} \rightarrow 0$ and  $1/2 > \Omega^{(M)} > 0$ for $\omega < 0$ and $\omega = 0$, respectively (we have not considered $\omega > 0$). The models studied in section $6.3$ illustrate the second case. 

In this work  we have not discussed the physical mechanism leading to   DE  models that cross the $W = - 1$ barrier. This is an important and interesting question  but this is out of the scope of the present work.

\end{document}